\documentclass[11pt]{article}
\usepackage{graphicx}
\usepackage{epsfig}
\usepackage{hyperref}
\usepackage[usenames,dvipsnames]{color}
\usepackage{wrapfig}
\usepackage{amsmath}
\def\lsim{\raise0.3ex\hbox{$<$\kern-0.75em\raise-1.1ex\hbox{$\sim$}}}
\def\gsim{\raise0.3ex\hbox{$>$\kern-0.75em\raise-1.1ex\hbox{$\sim$}}}

\def\mean#1{\left<#1\right>}
\def\Journal#1#2#3#4{{#1}{\bf #2} (#4) #3}

\def\EPJC{{Eur. Phys. J. C}}

\def\NCA{Nuovo Cimento\ }

\def\NPB{{Nucl. Phys. B}}
\def\PLB{{Phys. Lett. B}}

\def\PRL{Phys. Rev. Lett.\ }
\def\PRD{{Phys. Rev. D}}
\def\PRC{{Phys. Rev. C}}
\def\PR{Phys. Rev.\ }

\def\ARNPS{{Ann. Rev. Nucl. Part. Sci.\ }}

\normalsize
\begin{document}
\title{Latest Results from BNL and RHIC-2013}
\author{M.~J.~Tannenbaum 
\thanks{Research supported by U.S. Department of Energy, DE-AC02-98CH10886.}
\\ Physics Department, 510c,\\
Brookhaven National Laboratory,\\
Upton, NY 11973-5000, USA\\
mjt@bnl.gov}\maketitle
\maketitle
\thispagestyle{empty}
\section{Introduction}\label{sec:introduction}
\begin{figure}[!t]
\begin{center}
\includegraphics[width=0.61\textwidth]{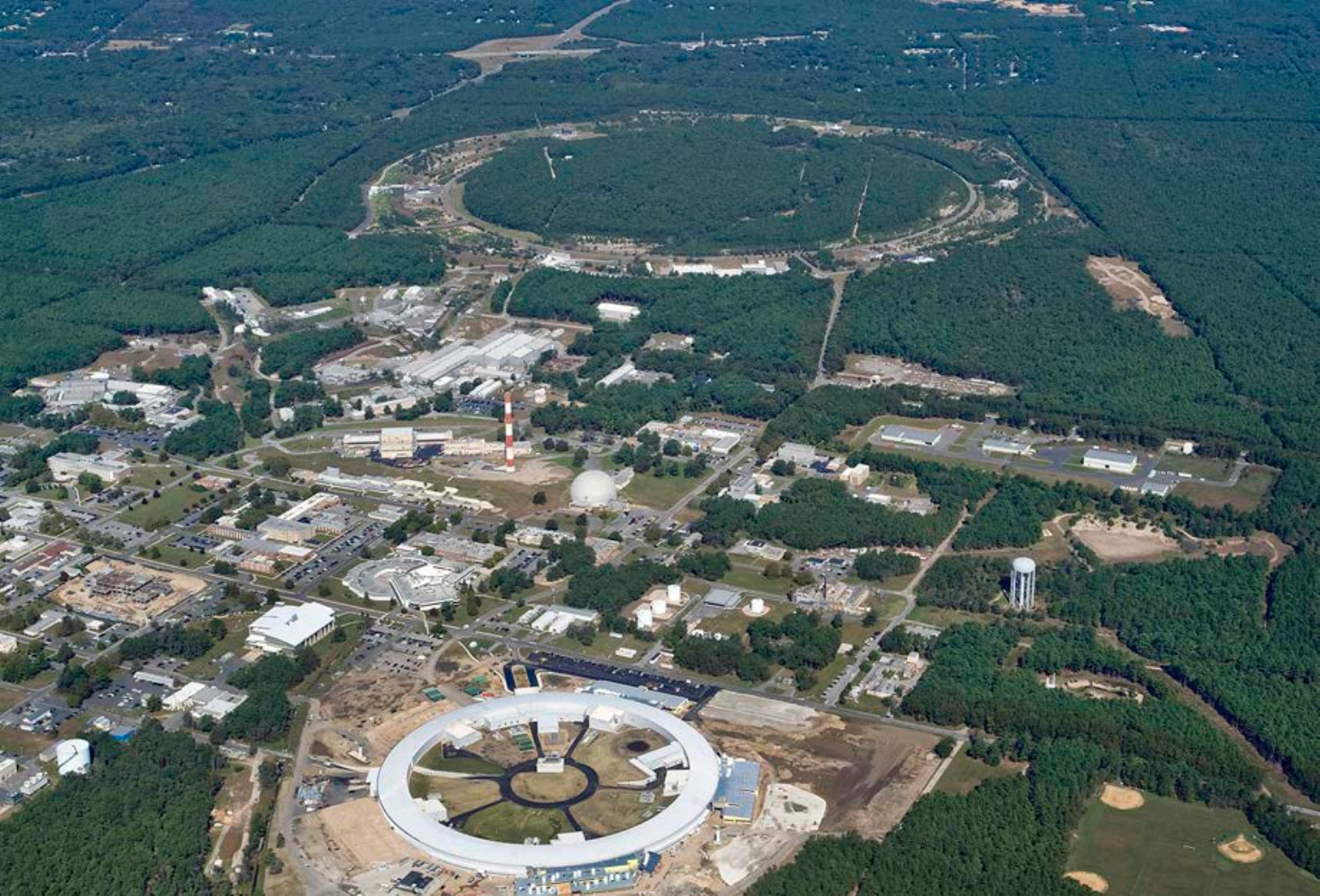}
\includegraphics[width=0.375\textwidth]{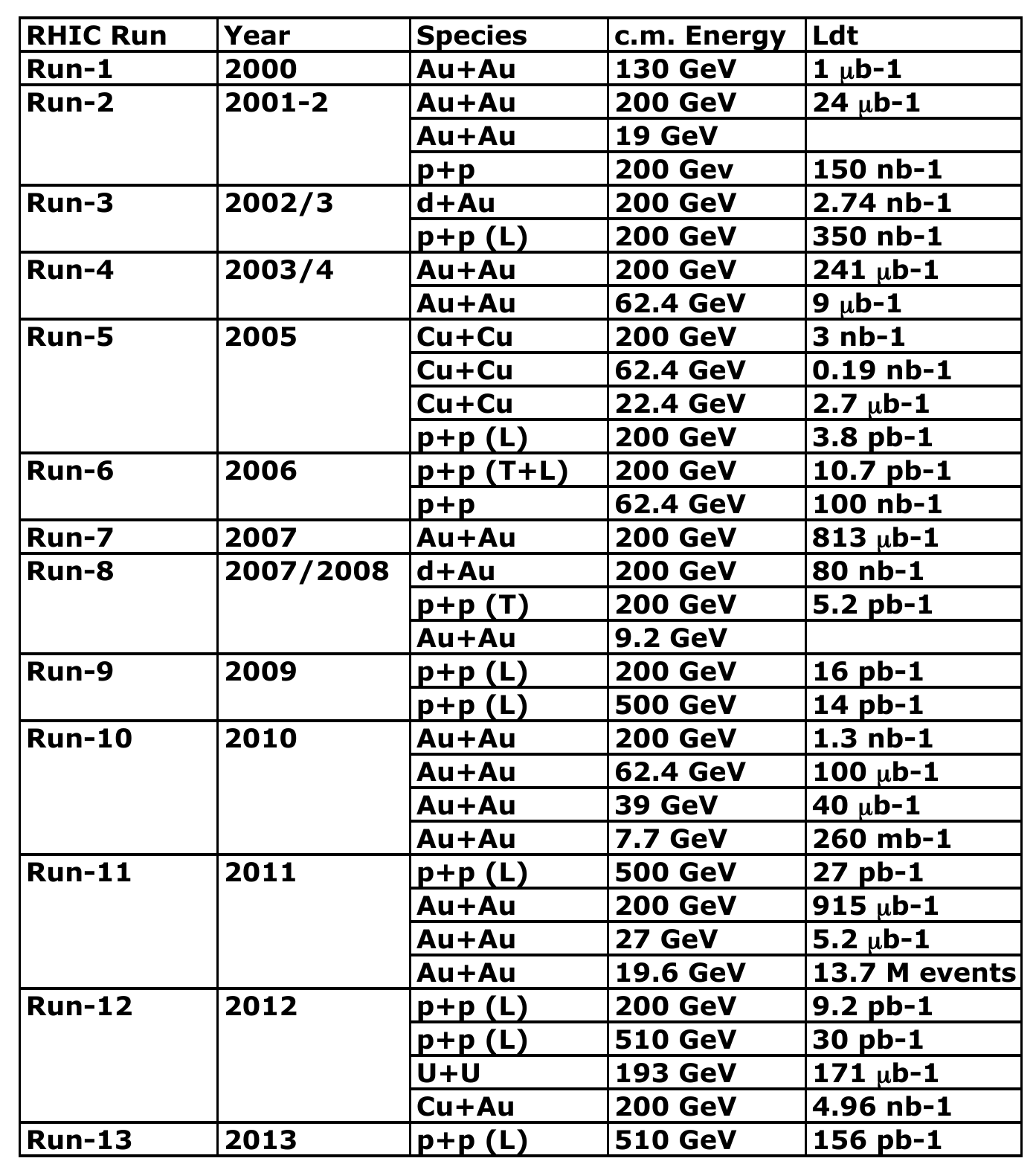}
\end{center}\vspace*{-1.5pc}
\caption[]{a)(left) Aerial view of BNL. In addition to RHIC, at the top, there are two synchrotron light sources visible, the rings at the bottom. The dome in the center is the decommissioned High Flux Beam Reactor (1965-1996); and the chimney is the last remnant of the original (1950-1968) Graphite Research Reactor. b) Year, species, $\sqrt{s_{NN}}$, integrated luminosity and proton polarization  ({\bf L}ongitudinal, {\bf T}ransverse ), for RHIC runs.}
\label{fig:BNLRHICoverview}\vspace*{-0.5pc}
\end{figure}

Quite different from CERN, Fermilab and SLAC, Brookhaven National Laboratory (BNL) is a multipurpose laboratory, founded in 1947 to promote basic research in the physical, chemical, biological and engineering aspects of the atomic sciences and for the purpose of the design, construction and operation of large scientific machines that individual institutions could not afford to develop on their own. The present mission has expanded to both basic and applied research at the frontiers of science, including nuclear and high-energy physics; physics and chemistry of materials; nanoscience; energy and environmental research; national security and nonproliferation; neurosciences; structural biology; computational sciences; and to provide ``cutting edge'' research facilities for these purposes. BNL is located on Long Island roughly 100km east of New York City. In fact the Relativistic Heavy Ion Collider (RHIC) at BNL can be seen from outer space since it is not buried in a tunnel but is in an enclosure on the surface, which is covered by earth for shielding (Fig.~\ref{fig:BNLRHICoverview}a).  

RHIC is composed of two independent rings, of circumference 3.8 km, containing a total of 1,740 superconducting magnets. RHIC can collide any species with any other species and since beginning operation in the year 2000 has provided collisions at 13 different values of nucleon-nucleon c.m. energy, $\sqrt{s_{NN}}$, and nine different species combinations (counting differently polarized protons as different species). The performance history of RHIC with A+A and polarized p-p collisions is shown in Fig.~\ref{fig:BNLRHICoverview}b. 

\section{News from BNL since ISSP2012} In keeping with the spirit of the title, I review the latest news from BNL before going to the RHIC physics results. In September 2012, the panel appointed by the U.~S.~Department of Energy (DoE), led by Prof. Robert Tribble, to decide whether RHIC, Jefferson Laboratory  in Newport News, VA, or the Facility for Rare Isotope Beams (FRIB) project at Michigan State University should be closed (due to insufficient funding) concluded with a good outcome. They recommended that a 2\% real increase per year in the DoE nuclear physics budget would save all three labs. This was included in the President's 2014 budget.\footnote{However, the Fiscal Year 2014 started on October 1, 2013, with the U.~S.~Government shut down for 16 days due to the lack of an approved budget, so \ldots}  

There were several important personnel changes at BNL. On January 8, 2013, Berndt Mueller, a leading Nuclear Physics Theorist and Administrator from Duke University became the new Associate Laboratory Director responsible for the Nuclear and Particle Physics programs at BNL. In March 2013, Doon Gibbs, who had been the Deputy Laboratory Director, became the new Laboratory Director, while the previous Laboratory Director, Sam Aronson, became the Director of the RIKEN-Brookhaven Research Center at BNL, replacing Nick Samios. About the same time, a new world-class interdisciplinary Science Laboratory building was completed with 60 standard laboratories and 4 specialized laboratories: a humidity controlled dry room, two ultra-low vibration laboratories, and a laboratory which connects the Molecular Beam Epitaxy (MBE) machine, which builds custom designed thin films one atomic layer at a time, to one of the ultra-low vibration labs via a vacuum-locked system which allows the MBE-created samples to be transported for analysis 
without exposing them to air. Then, a week later, 
the U.~S.~Department of Energy, the owner of BNL, announced that they would initiate a competetion for a new contractor to Manage \& Operate BNL for the DoE, with evident dissatisfaction in the performance of the present contractor who had managed BNL since 1998, having replaced the original founders of the laboratory, a consortium of nine major northeastern research universities, who had operated the laboratory from 1947-1998.

In June, the 15 m diameter precision storage ring from the BNL muon \mbox{$g-2$} experiment~\cite{BNL-g-2} began a circuitous (5000 km) very delicate cross-contry trip to Fermilab (only 1440 km in a straight line) involving custom built trucks and a specially prepared barge which brought the magnet down the East Coast, around the tip of Florida and up the Mississippi River to Illinois. This new muon $g-2$ experiment at Fermilab would be the fifth such experiment, which was pioneered at CERN. 

\begin{figure}[!h]
\begin{center}
\includegraphics[width=0.375\textwidth]{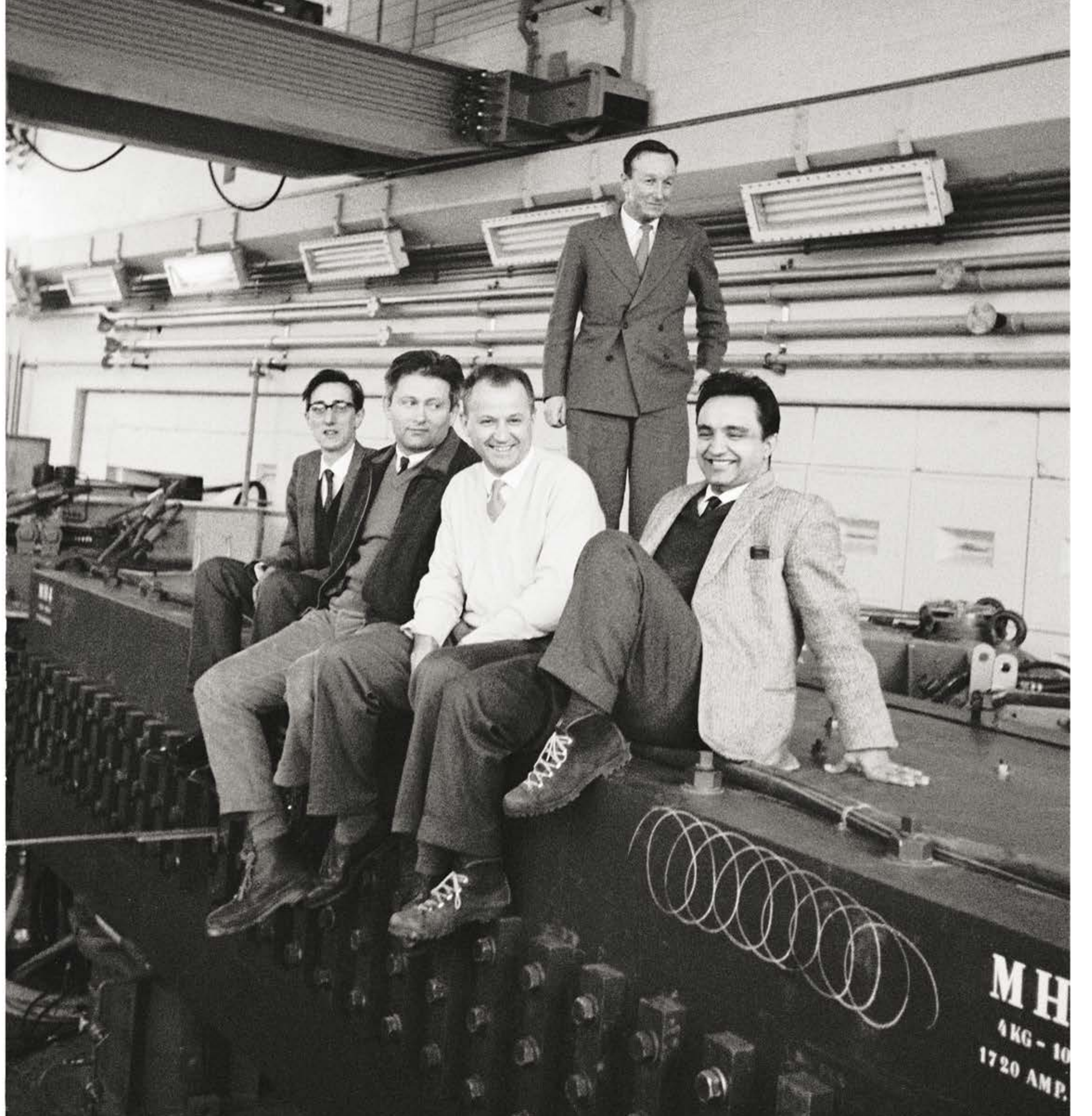}
\includegraphics[width=0.61\textwidth]{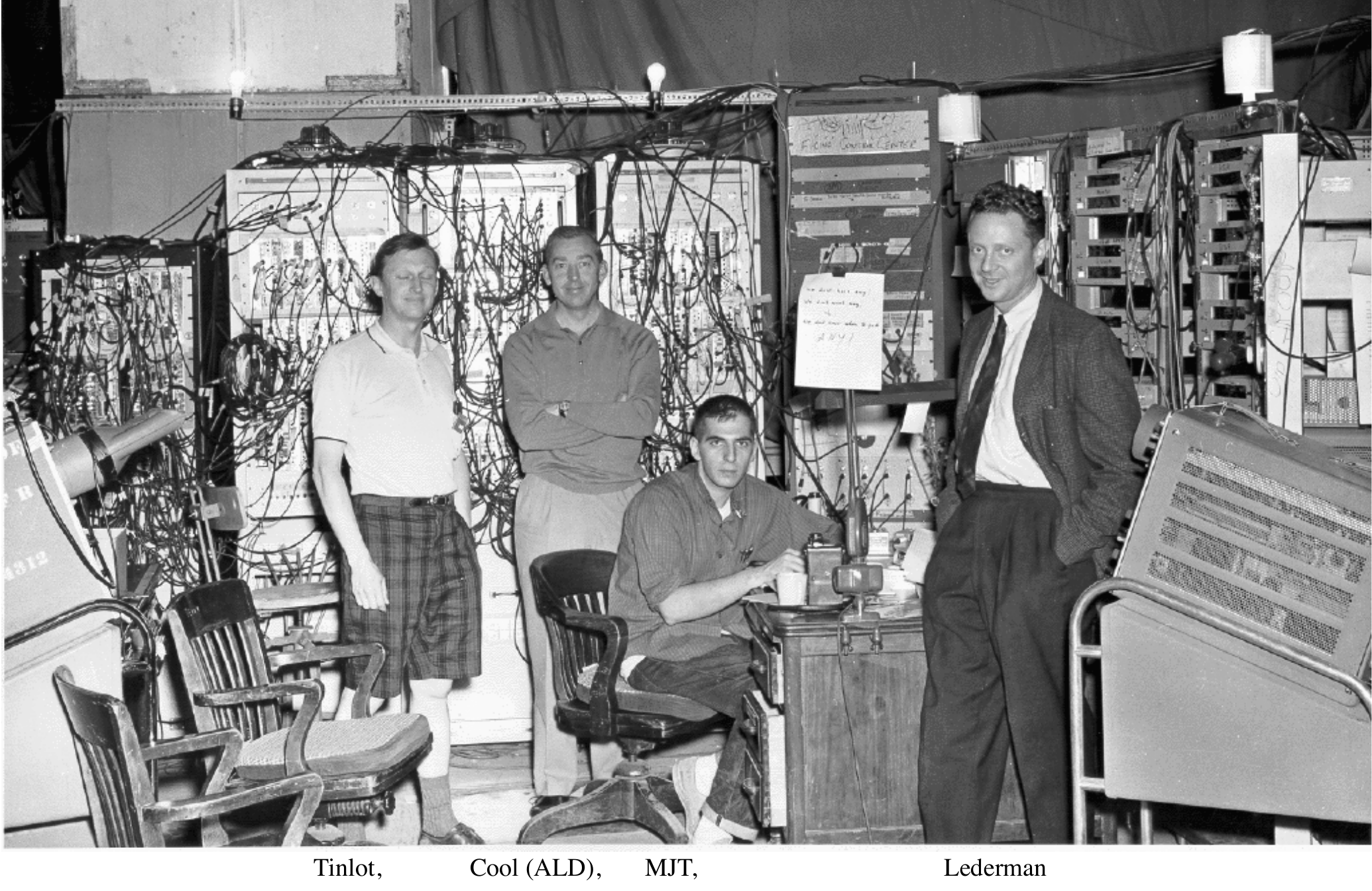}
\end{center}\vspace*{-1.5pc}
\caption[]{a)(left) First $g-2$ experiment, CERN, c. 1959. Left-to-Right: Sens, Charpak, Muller, Farley, Zichichi. b) (right) $\mu-p$ experiment BNL, c. 1963: L-R: Tinlot, Cool, Tannenbaum, Lederman.} 
\label{fig:NinoMike62}
\end{figure}

By some incredible coincidence, a few weeks before this ISSP2013 school, the June 2013 CERN Courier reprinted an article from 1970 with the title ``Preparing for a third `$g-2$'.'' 
This brought back good memories to me because I had worked on the second $g-2$ experiment when I was a post-doc at CERN in 1965-66~\cite{g-2-2-1966}; but, of course, Prof. Zichichi worked on the ground-breaking original $g-2$ experiment at CERN in 1959-1961~\cite{CERN-first-g-2-final}. Figure~\ref{fig:NinoMike62}a shows the original $g-2$ team seated on top of their 6 m magnet, including a $\sim30$ year old Nino. Fig.~\ref{fig:NinoMike62}b was taken in 1963 on the BNL-AGS floor during my thesis experiment where the three Professors were discussing what the $\sim24$ year old graduate-student should do next. My thesis was muon-proton elastic scattering~\cite{MuP1968} to find out ``Why does the muon weigh heavy?'' Even today, with Prof. Higgs in the audience, we still don't know! Other experiments at that time did better: the ``two-neutrino experiment''~\cite{Danby1962} was in the beam to the left (Nobel Prize); while on the right, over the AGS machine, in ``inner Mongolia'', CP violation~\cite{Fitch1963} was discovered (Nobel Prize). Those were the days; but even more excitement lay ahead. 

\section{ICHEP1972: Hard-Scattering, Quarks, and QCD} At the International Conference on High Energy Physics (ICHEP) in 1972, there were three momentous developments that inform our work today: 
\begin{itemize}
\item The discovery in p-p collisions at the CERN ISR of production of particles with large transverse momentum ($p_T$) which proved that the partons of Deeply Inelastic Scattering (DIS) interacted with each other much more strongly than electromagnetically. 
\item Measurements of DIS in neutrino scattering presented by Don Perkins who proclaimed that ``In terms of constituent models, the fractionally charged (Gell-Mann/Zweig) quark model is the only one which fits both the electron and neutrino data.''
\item The origin of QCD in the presentation by Harald Fritzsch and Murray Gell-Mann with the title ``Current Algebra: Quarks and What Else?''
\end{itemize}

\begin{figure}[!h]
   \begin{center}
\includegraphics[width=0.48\textwidth]{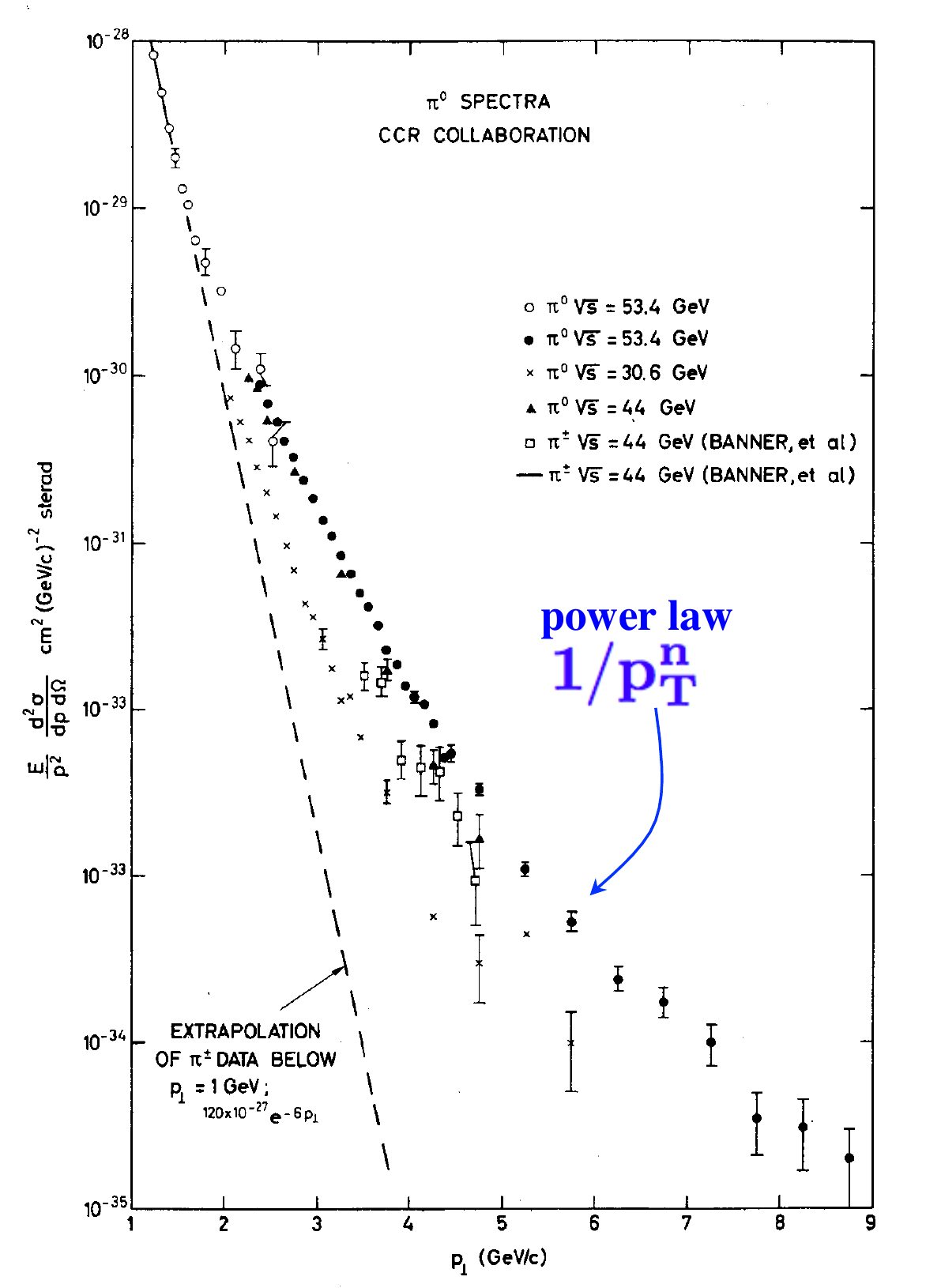}
\raisebox{0.2pc}{\includegraphics[width=0.51\textwidth]{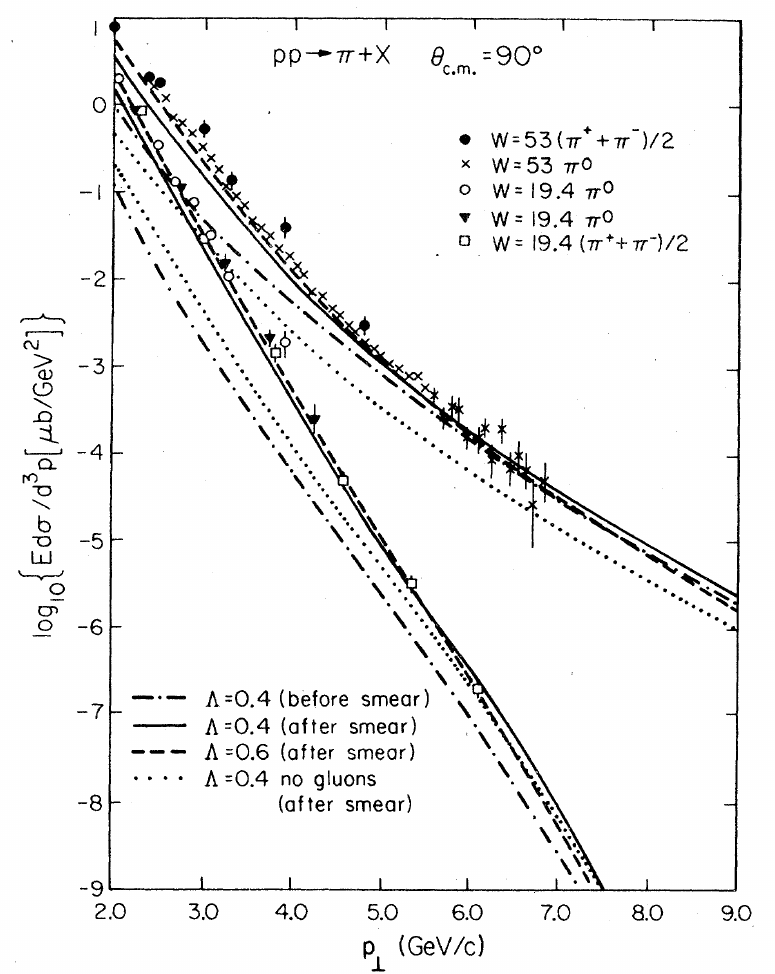}}
\end{center}\vspace*{-1.7pc}
\caption[]{a)(left) Plot of invariant single $\pi^0$ cross section vs. $p_T$ for several $\sqrt{s}$ from CCR at the CERN-ISR~\cite{CoolICHEP72}. b)(right) Feynman, Field, and Fox~\cite{FFF} QCD calculation of mid-rapidity high-$p_T$ $\pi$ spectra at $\sqrt{s}=19.4$ and 53 GeV, with and without $k_T$ smearing, for two values of $\Lambda_{\rm QCD}$.} 
\label{fig:QCDworks}\vspace*{-0.8pc}
\end{figure}

Figure~\ref{fig:QCDworks}a shows the first observation of scattering at large $p_T$~\cite{CoolICHEP72}. The $\pi^0$ spectrum breaks away from the $e^{-6p_T}$ dependence known since cosmic ray measurements, with a power-law spectrum that flattens as the c.m energy, $\sqrt{s}$, is increased. Excellent cooperation of experimentalists and theorists showed in 1978 that these data could be explained by QCD~\cite{OwensKimel,FFF} if the quarks in a nucleon had ``intrinsic'' transverse momentum, $k_T\approx 1$ GeV/c (Fig.~\ref{fig:QCDworks}b). 
Although these QCD calculations in agreement with the high $p_T$ single particle spectra were published in 1978, most experimentalists in the U.~S., notably at the first Snowmass conference in July 1982, were skeptical because of evidence against jets presented at the ICHEP1980 by a CERN experiment, NA5~\cite{NA5PLB112}. 

\begin{figure}[!h]
   \begin{center}
\includegraphics[width=0.48\textwidth,height=0.5\textwidth]{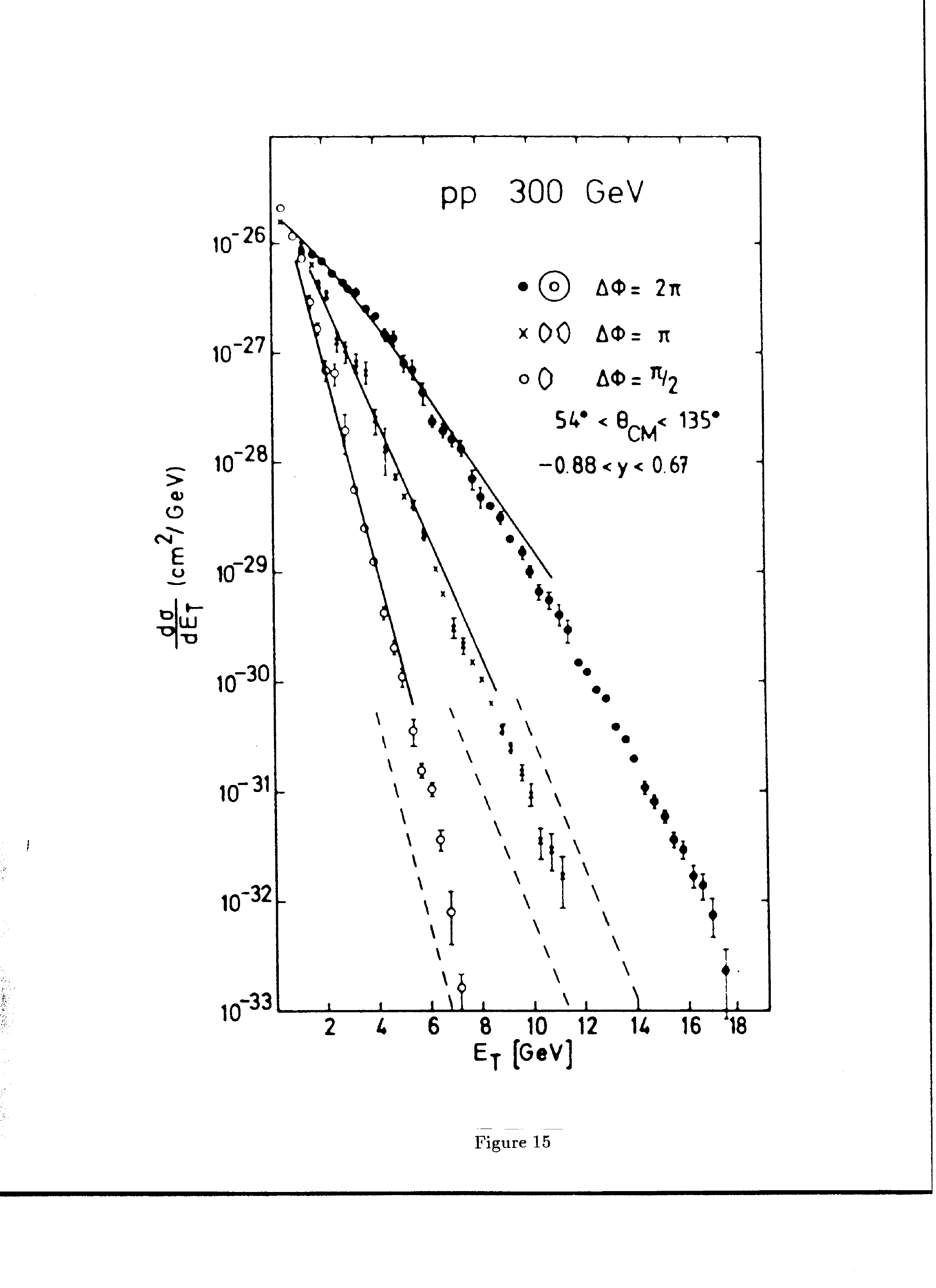}\hspace*{-1pc}
\raisebox{0.2pc}{\includegraphics[width=0.53\textwidth]{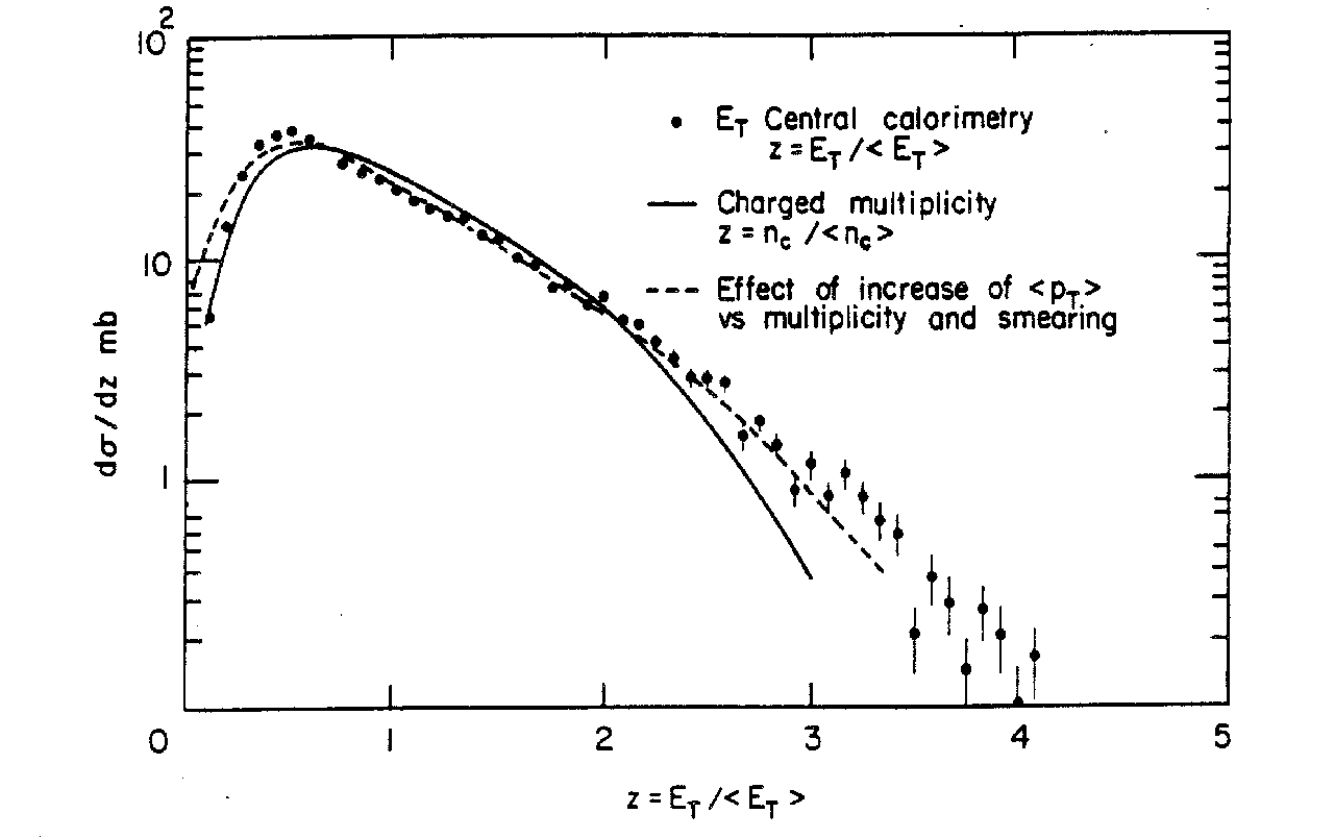}}
\end{center}\vspace*{-0.25in}
\caption[]{a)(left) $E_T$ distributions~\cite{NA5PLB112} in the solid angles indicated. Predictions from soft (low $p_T$) multiparticle production and QCD hard-scattering are shown by solid and dashed curves respectively. b) (right) $E_T$ distribution~\cite{UA1Paris82} for $|\eta|\leq3$ from $\bar{\rm p}$-p collisions at $\sqrt{s}=540$ GeV in the KNO type variable, $z=E_T/\mean{E_T}$, used for multiplicity~\cite{UA1PLB107}. } 
\label{fig:firstET}\vspace*{-1pc}
\end{figure}

Bjorken had proposed in 1973~\cite{BjPRD8} that jets from the fragmentation of high $p_T$ scattered partons should be observed using ``$4\pi$'' hadron calorimeters.  The first large aperture measurement was by NA5~\cite{NA5PLB112} (Fig.~\ref{fig:firstET}a) who showed a transverse energy ($E_T$) spectrum at ICHEP1980, where the sum:\vspace*{-0.2pc}
      \begin{equation}
          E_T=\sum_i E_i\ \sin\theta_i
          \label{eq:ETdef}\vspace*{-0.3pc}
          \end{equation}
is taken over all particles emitted into a fixed solid angle for each event. In Fig.~\ref{fig:firstET}a~\cite{NA5PLB112}, 
the solid angle varies from full azimuthal acceptance, $\Delta\phi=2\pi$, in the c.m. rapidity range $-0.88<y<0.76$, to smaller azimuthal regions as shown on the figure. The striking results, which contradicted a previous claim from Fermilab~\cite{E260NPB134}, were: i) no jets were seen in the full azimuth data; ii) all the data were far above the QCD predictions; iii) the large $E_T$ observed was the result of ``a large number of particles with a rather small transverse momentum''. 

As we shall see below, $E_T$ distributions are very important in Relativistic Heavy Ion (RHI) physics since they can be used to characterize and study the nuclear geometry of an A+B reaction on an event-by-event basis. The strong relation between $E_T$ and multiplicity distributions and the absence of jets in these distributions was emphasized in a talk by UA1 at ICHEP 1982 (Fig.~\ref{fig:firstET}b)~\cite{UA1Paris82}. Ironically, this talk immediately followed a talk by UA2 which provided the first evidence for a di-jet from hard-scattering at a level 5-6 orders of magnitude down in the $E_T$ distribution at $\sqrt{s}=540$ GeV.

There are many additional important results in high-energy physics from this period that are relevant to both QCD and RHI physics, which must be skipped in this brief introduction. I have covered some of these results in previous ISSP lectures and proceedings; but this year, I wrote a book with Jan Rak~\cite{RATCUP} which covers this information in detail, which is the real introduction to what follows. 
\section{QGP Physics---RHIC Highlights 2013} 
The Quark Gluon Plasma (QGP) was discovered at RHIC, and announced on April 19, 2005, with properties of a hot, dense, inviscid liquid,  rather than the gas of free gluons and quarks expected.\footnote{This is discussed in more detail in my proceedings from the ISSP2011, also available as arXiv:1201.5900.} Although soft ($p_T\lsim 2$ GeV/c) physics dominates particle production in p-p, p+A and A+A collisions and presumably in the thermalized QGP, with $T\approx 300-600$ MeV as estimated from ``thermal photons'' measured in the range $1\leq p_T\leq 3$ GeV/c~\cite{AdarePRL104},  the use of hard-scattering as an in-situ probe of the medium in RHI collisions was an important and productive  innovation at RHIC. The effect of the medium on outgoing hard-scattered partons produced by the    \centerline{\mbox{\includegraphics[width=0.9\textwidth]{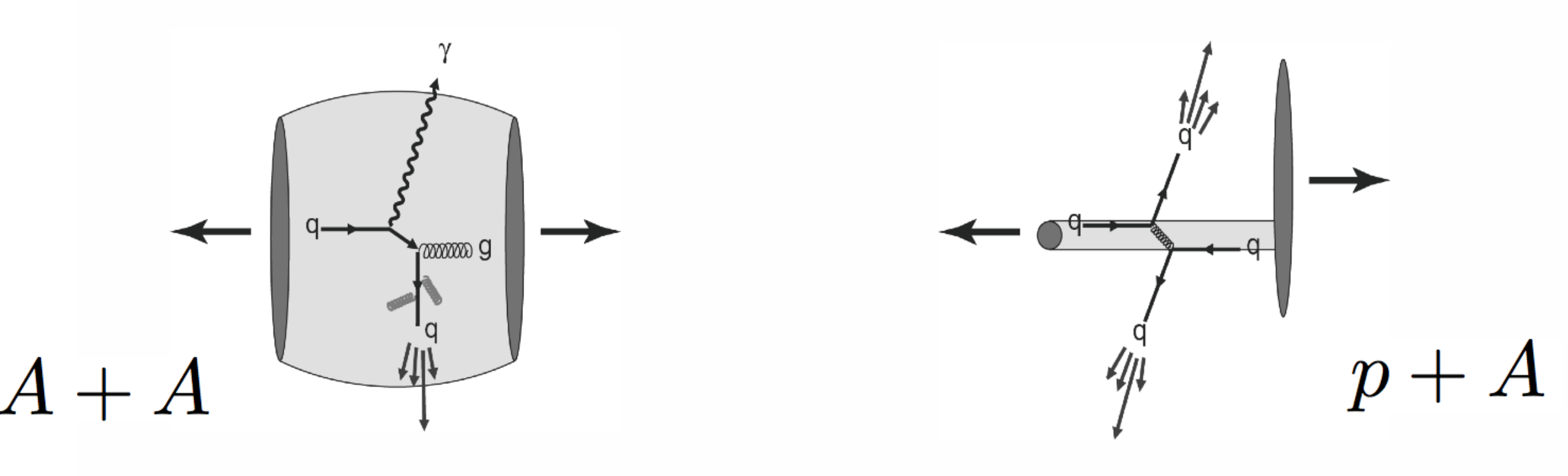} }} 
\mbox{initial}  A+A collision is determined by comparison to measurements in p+A (or d+A) collisions, where no (or negligible) medium is produced. The impact parameter of the nuclear collision, the distance between the centers of the colliding nuclei, is measured by the upper percentile (e.g. top 5\%) of an $E_T$ or multiplicity distribution since the multiplicity increases with increasing overlap of the nuclei (centrality).  The highlights this year include hard-scattering and soft physics results in Au+Au together with some important new results in d+Au.

\begin{figure}[!t]
\includegraphics[height=0.26\textheight]{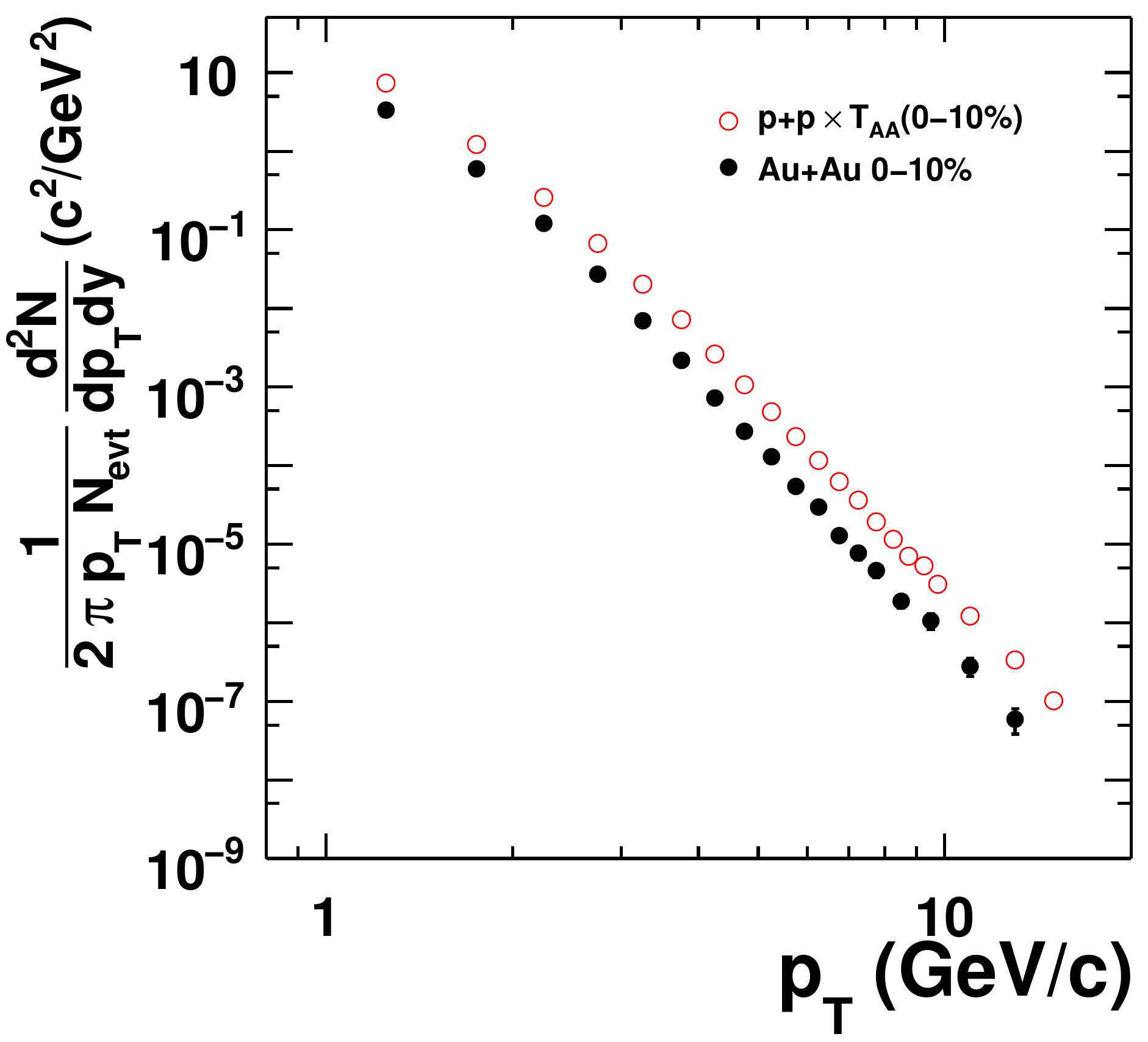}
\hspace*{0.001\textwidth} \includegraphics[height=0.267\textheight]{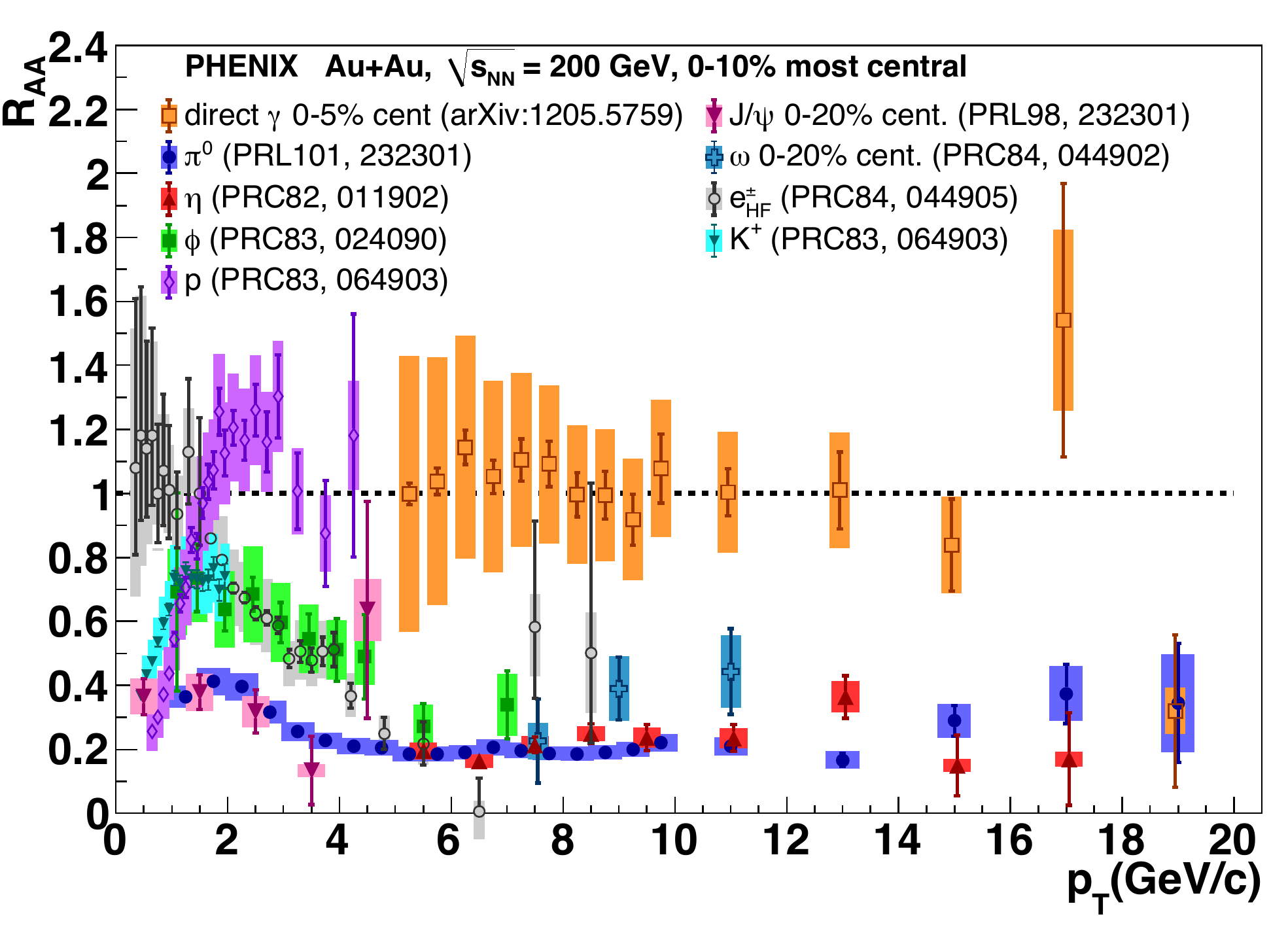}\vspace*{-0.5pc}
\caption{a) (left) Log-log plot of invariant yield of $\pi^0$ at $\sqrt{s_{NN}}=200$ GeV as a function of transverse momentum $p_T$ in p-p collisions multiplied by $\mean{T_{AA}}$ for Au+Au central (0--10\%) collisions compared to the Au+Au measurement~\cite{ppg054}. b) (right) $R_{AA}(p_T)$ for all identified particles so far measured by PHENIX in Au+Au central collisions at $\sqrt{s_{NN}}=200$ GeV.}
\label{fig:Tshirt}\vspace*{-0.5pc}
\end{figure}

The suppression of high $p_T$ particles in Au+Au collisions at RHIC is reviewed in Fig.~\ref{fig:Tshirt}. For $\pi^0$ (Fig.~\ref{fig:Tshirt}a)~\cite{ppg054} the hard-scattering in p-p collisions is indicated by the power law behavior $p_T^{-n}$ for the invariant cross section, $E d^3\sigma/dp^3$, with $n=8.1\pm 0.05$ for $p_T\geq 3$ GeV/c at $\sqrt{s_{NN}}=200$ GeV.  The Au+Au data at a given $p_T$ can be characterized either as shifted lower in $p_T$ by $\delta p_T$ from the point-like scaled p-p data at $p'_T=p_T+\delta p_T$, or shifted down in magnitude, i.e. suppressed. In Fig.~\ref{fig:Tshirt}b, the suppression of the many identified particles measured by PHENIX at RHIC is presented as the Nuclear Modification Factor, 
$R_{AA}(p_T)$, the ratio of the yield of e.g. $\pi$ per central Au+Au collision (upper 10\%-ile of observed multiplicity)  to the point-like-scaled p-p cross section, where $\mean{T_{AA}}$ is the average overlap integral of the nuclear thickness functions: 
   \begin{equation}
  R_{AA}(p_T)=\frac{{d^2N^{\pi}_{AA}/dp_T dy N_{AA}}} { \mean{T_{AA}} d^2\sigma^{\pi}_{pp}/dp_T dy} \quad . 
  \label{eq:RAA}
  \end{equation}

The striking differences of $R_{AA}(p_T)$ in central Au+Au collisions for the many particles measured by PHENIX  (Fig.~\ref{fig:Tshirt}b) illustrates the importance of particle identification for understanding the physics of the medium produced at RHIC. Most notable are: the equal suppression of $\pi^0$ and $\eta$ mesons by a constant factor of 5 ($R_{AA}=0.2$) for $4\leq p_T \leq 15$ GeV/c, with suggestion of an increase in $R_{AA}$ for $p_T > 15$ GeV/c; the equality of suppression of $\pi^0$ and direct-single $e^{\pm}$ (from heavy quark ($c$, $b$) decay) at $p_T\gsim 5$ GeV/c; the enhancement of the protons for $2<p_T<4$ GeV/c (baryon anomaly). For $p_T\gsim 4$ GeV/c, the hard-scattering region,  the fact that all measured hadrons are suppressed, but direct-$\gamma$ are not suppressed, indicates that suppression is a medium effect on outgoing color-charged partons likely due to energy loss by coherent Landau-Pomeranchuk-Migdal radiation of gluons, predicted in pQCD~\cite{BDMPS}, which is sensitive to properties of the medium. 

\begin{figure}[!h]
   \begin{center}
\includegraphics[width=0.49\textwidth,height=0.38\textwidth]{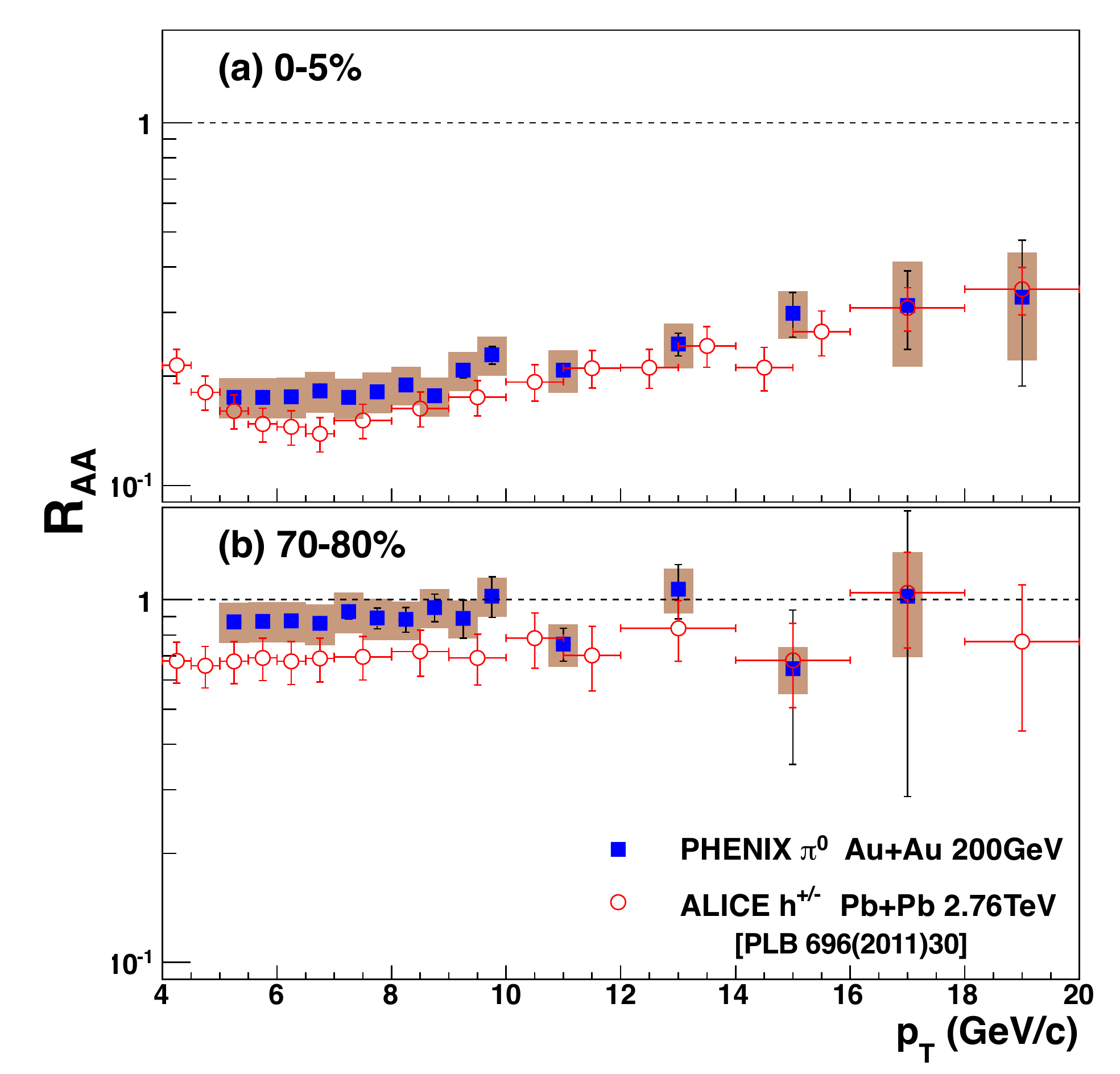}
\includegraphics[width=0.49\textwidth]{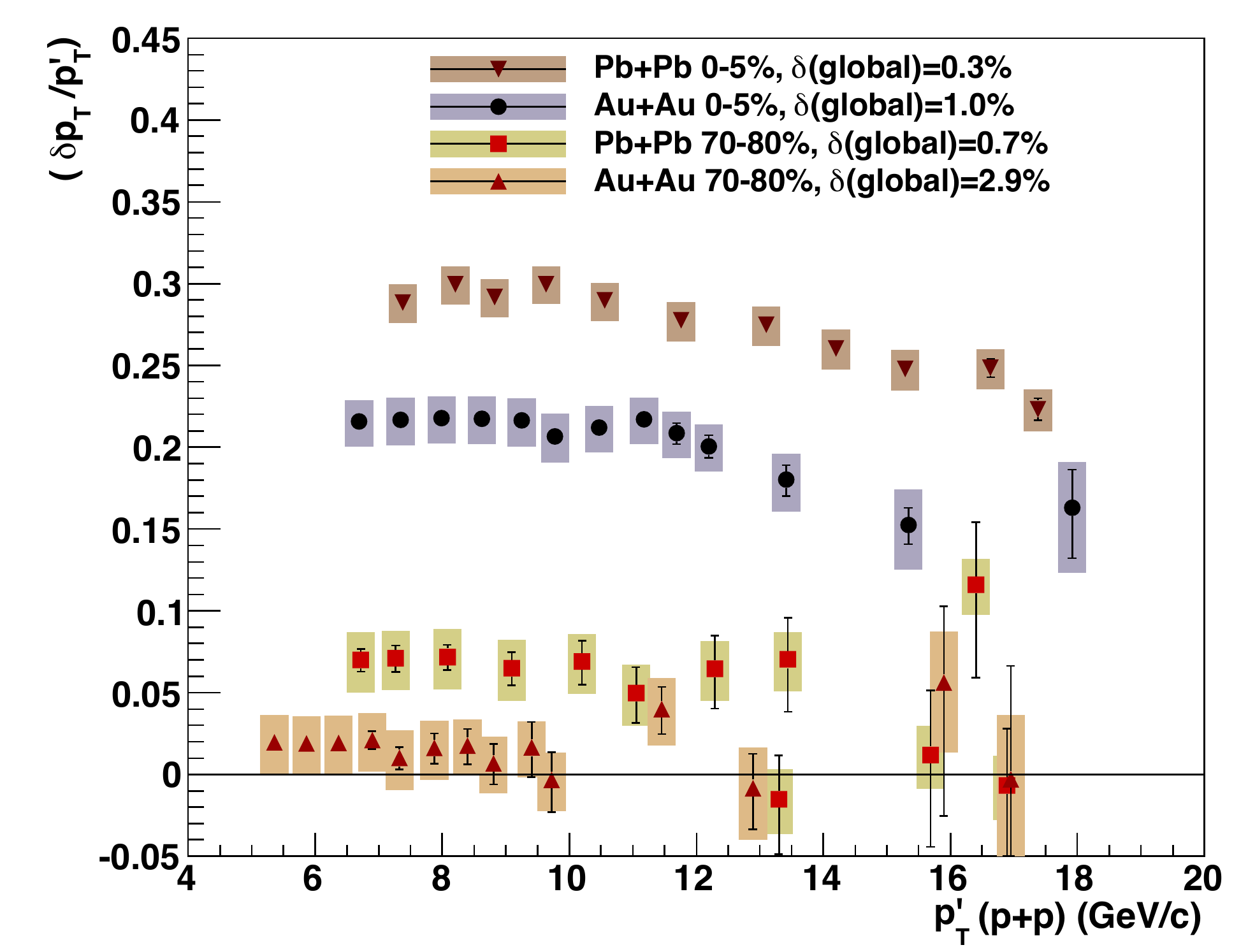}
\end{center}\vspace*{-1.7pc}
\caption[]{a) (left) $R_{AA}$ of $\pi^0$ in $\sqrt{s_{NN}}=200$ GeV central (0-5\%) and peripheral (70-80\%) Au+Au collisions~\cite{ppg133} at RHIC compared to non-identified charged hadron (${\rm h}^{\pm}$) $R_{AA}$ in $\sqrt{s_{NN}}=2.76$ TeV Pb+Pb collisions at LHC. b) (right) Fractional shift of $p_T$ spectrum $\delta p_T/p'_T$ vs. $p'_T$ (p-p) calculated by PHENIX~\cite{ppg133} for RHIC and LHC. 
\label{fig:ppg133}}\vspace*{-0.8pc}
\end{figure}

\begin{figure}[!b]
   \begin{center}
\includegraphics[width=0.99\textwidth]{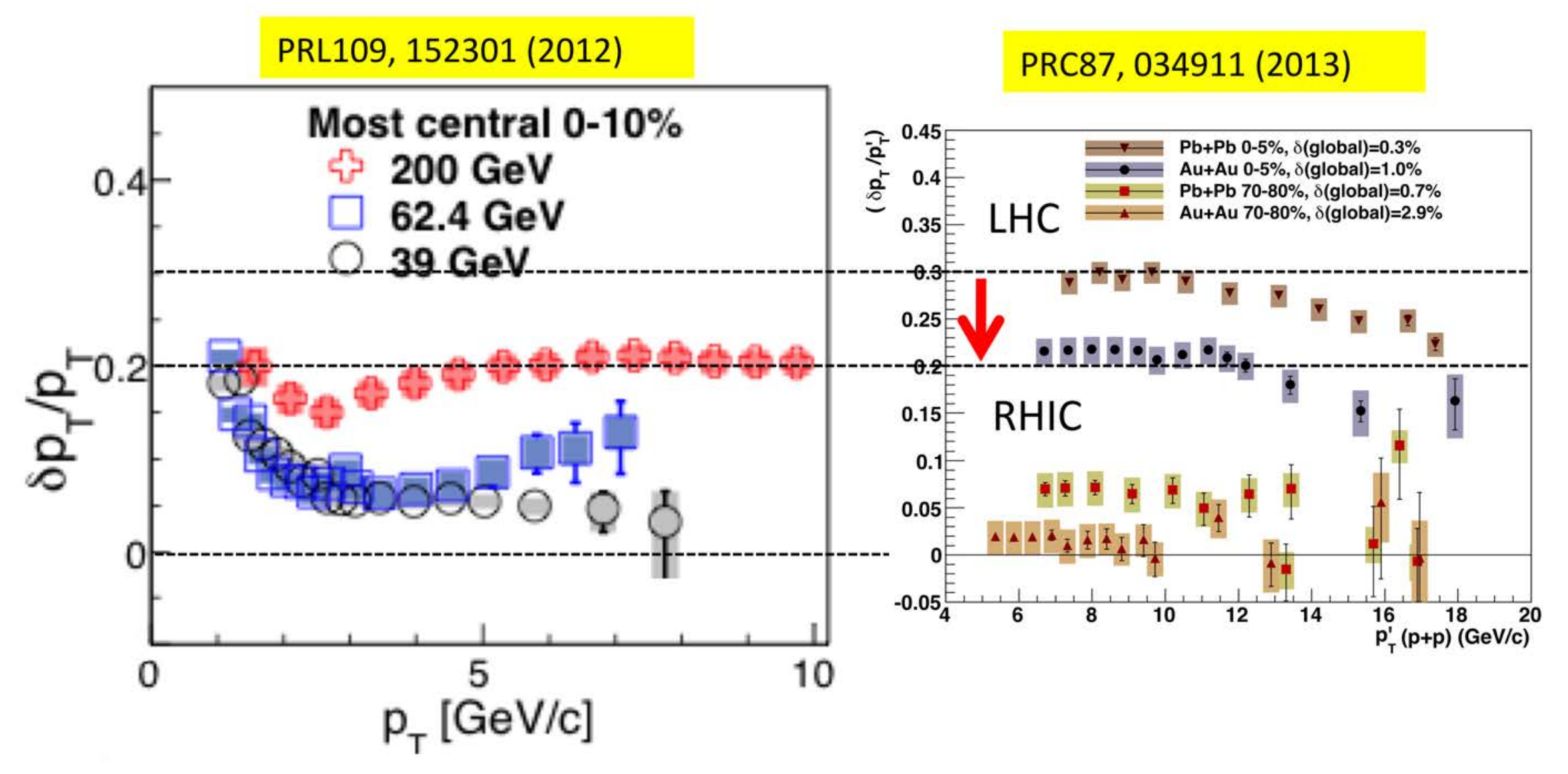}
\end{center}\vspace*{-1.7pc}
\caption[]{Fractional shift of $p_T$ spectrum $\delta p_T/p'_T$ in central A+A collisions from $\sqrt{s_{NN}}=39$~GeV to 2.76 TeV 
\label{fig:shiftallRHICLHC}}\vspace*{-1.8pc}
\end{figure}

Comparisons of the suppression of non-identified hadrons in $\sqrt{s_{NN}}=2.76$ TeV Pb+Pb collisions at LHC to the RHIC Au+Au $\pi^0$ data are also very interesting. This is shown both in terms of the suppression, $R_{AA}(p_T)$ (Fig.~\ref{fig:ppg133}a), and the fractional shift in the $p_T$ spectrum $\delta{p_T}/p'_T$ (Fig.~\ref{fig:ppg133}b).
Interestingly, despite more than a factor of 20 higher c.m. energy, the ALICE $R_{AA}$ data from LHC~\cite{ALICEPLB696} are nearly identical to the RHIC measurement~\cite{ppg133} for $5< p_T <20$ GeV/c. Since the exponent of the power-law at LHC ($n\approx 6$) is flatter than at RHIC ($n\approx 8$), a $\sim 40$\% larger  shift $\delta p_T/p'_T$ in the spectrum from p-p to A+A is required at LHC (Fig.~\ref{fig:ppg133}b) to get the same $R_{AA}$, which likely indicates $\sim 40$\% larger fractional energy loss at LHC in this $p_T$ range due to the probably hotter and denser medium. These measurements can be combined with the previous measurements at RHIC for $\sqrt{s_{NN}}=39$ and 62.4 GeV~\cite{ppg138} (Fig.~\ref{fig:shiftallRHICLHC}) to reveal a systematic increase of $\delta p_T/p'_T$ in central A+A collisions at $p'_T=7$ GeV/c, going from 5\% to 30\% over the c.m. energy range $\sqrt{s_{NN}}=39$~GeV to 2.76 TeV.   

This year, the major event was the p+Pb run at LHC which also spurred new or improved d+Au results from RHIC. Apart from one hard-scattering result to be presented first, all the results involve the predominant soft physics of multiplicity and $E_T$ distributions as well as flow. 
A new measurement of identified hadron production in both Au+Au d+Au at $\sqrt{s_{NN}}=200$ GeV~\cite{PXPRC88} gives  some insight into the baryon anomaly. 
         \begin{figure}[!h]
   \begin{center}
\includegraphics[width=0.49\textwidth]{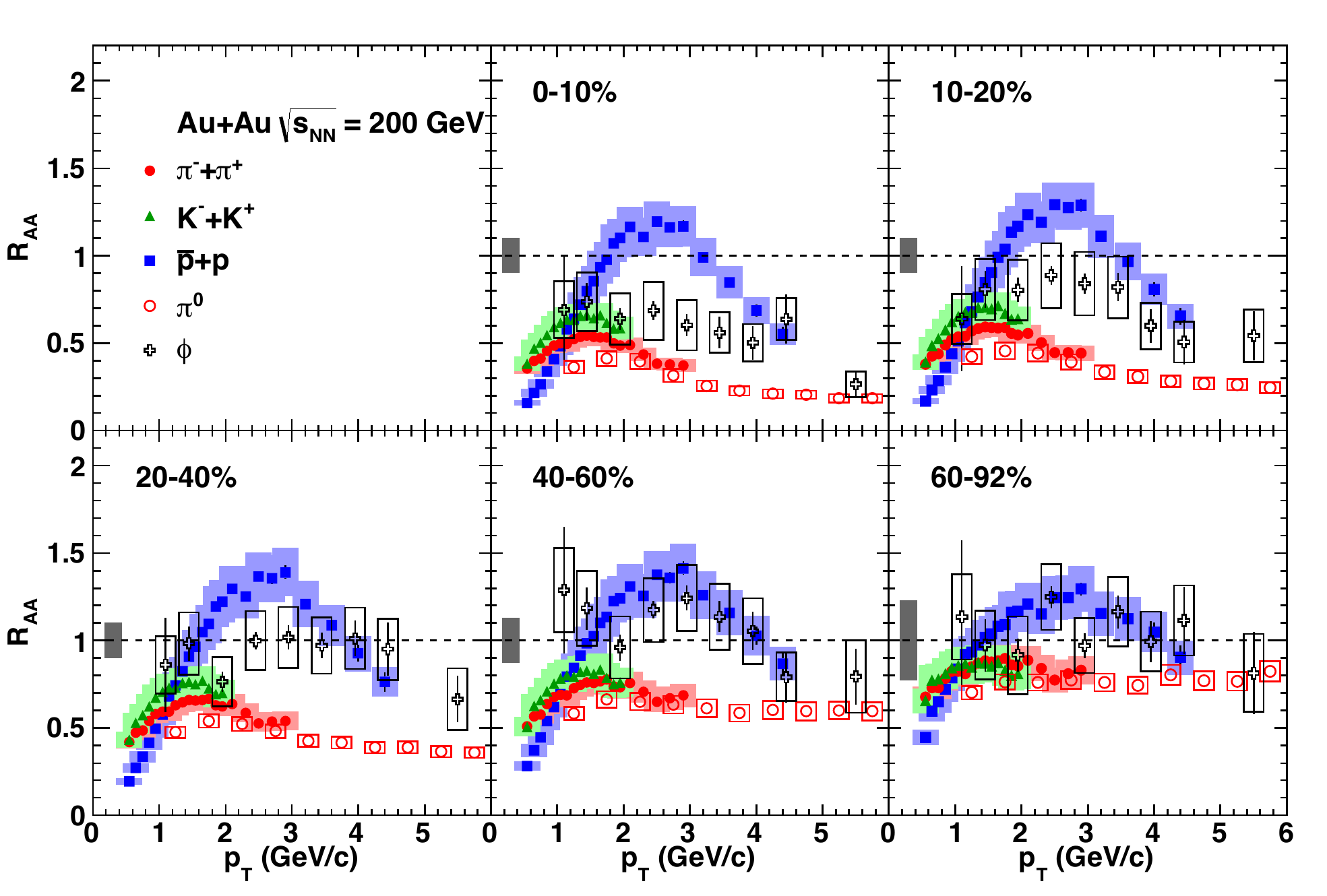}
\includegraphics[width=0.49\textwidth]{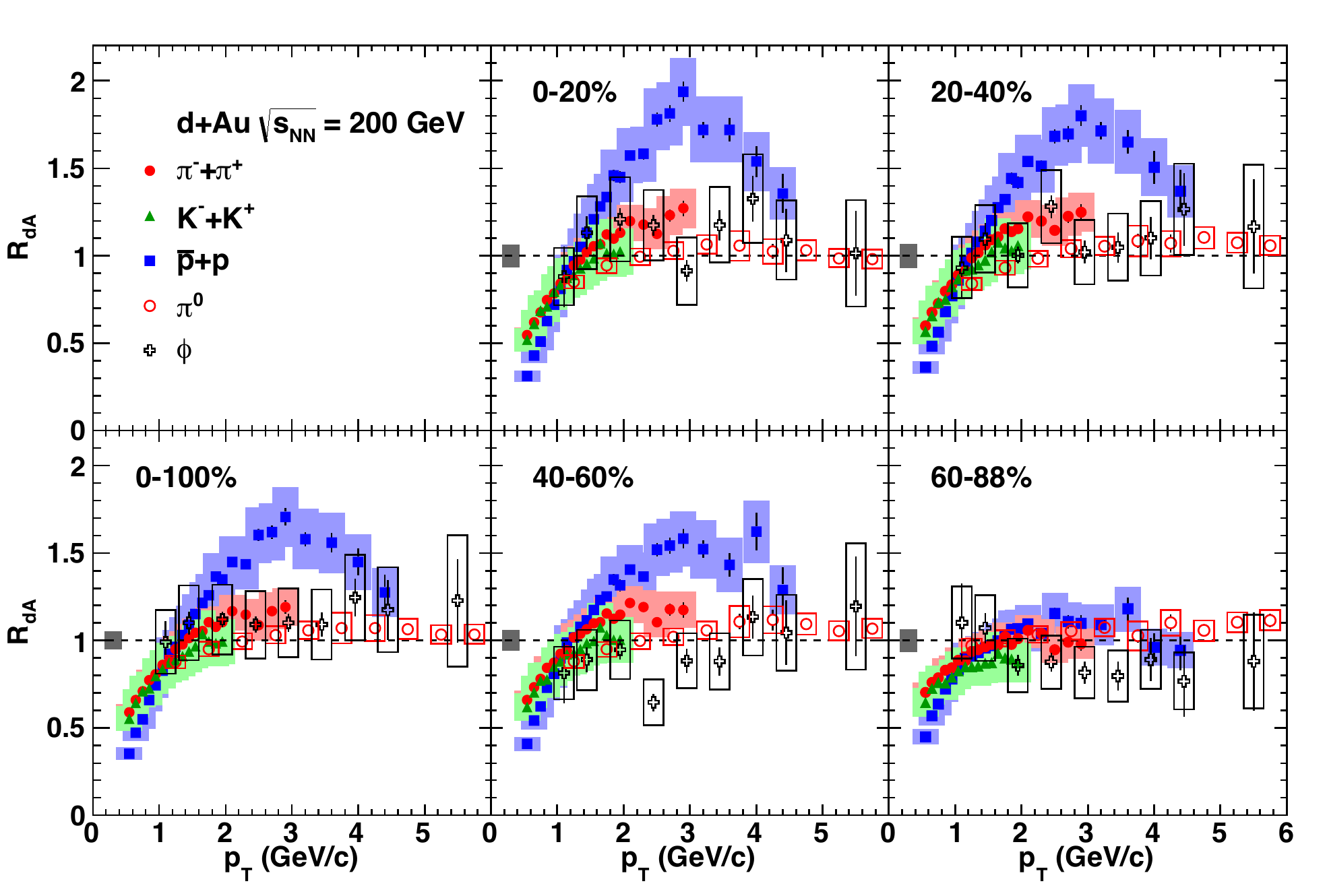}
\end{center}\vspace*{-1.7pc}
\caption[]{Measurements of $R_{AA}$ of identified particles as a function $p_T$ and centrality at $\sqrt{s_{NN}}=200$ GeV~\cite{PXPRC88}: a)(left) Au+Au; b) (right) d+Au. 
\label{fig:RAAdAuAuAu}}\vspace*{-0.1pc}
\end{figure}
Figure~\ref{fig:RAAdAuAuAu}a shows $R_{AA}$ in Au+Au for protons and mesons in the range $0.5<p_T<6.0$ GeV/c, where, in central collisions (0-10\%), all the mesons are suppressed for {$p_T>2$ GeV/c}  while the protons are enhanced for $2<p_T<4$ GeV/c and then become suppressed at larger $p_T$. The d+Au results in Fig.~\ref{fig:RAAdAuAuAu}b show no effect for the mesons, $R_{AA}\approx 1$ out to $p_T=6$ GeV/c; while the protons show a huge enhancement (Cronin effect) in all centralities except for the most peripheral (60-88\%).  This suggests the need for a common explanation of the proton enhancement in both Au+Au and d+Au collisions, which is lacking at present.  

\begin{figure}[!h]
   \begin{center}
\includegraphics[width=0.52\textwidth]{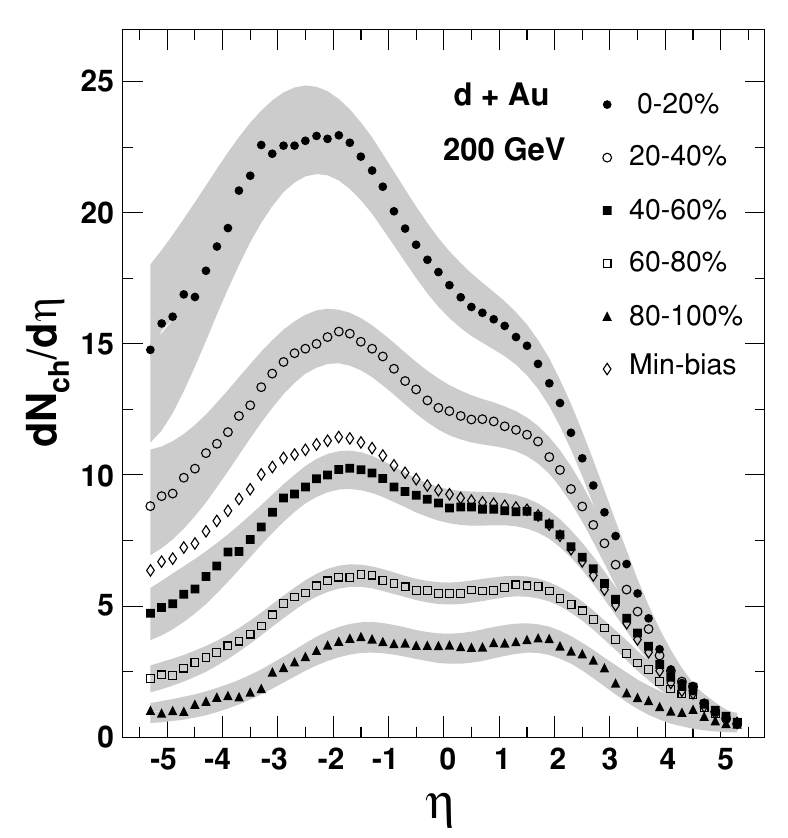}\hspace{0.1pc}
\includegraphics[width=0.46\textwidth]{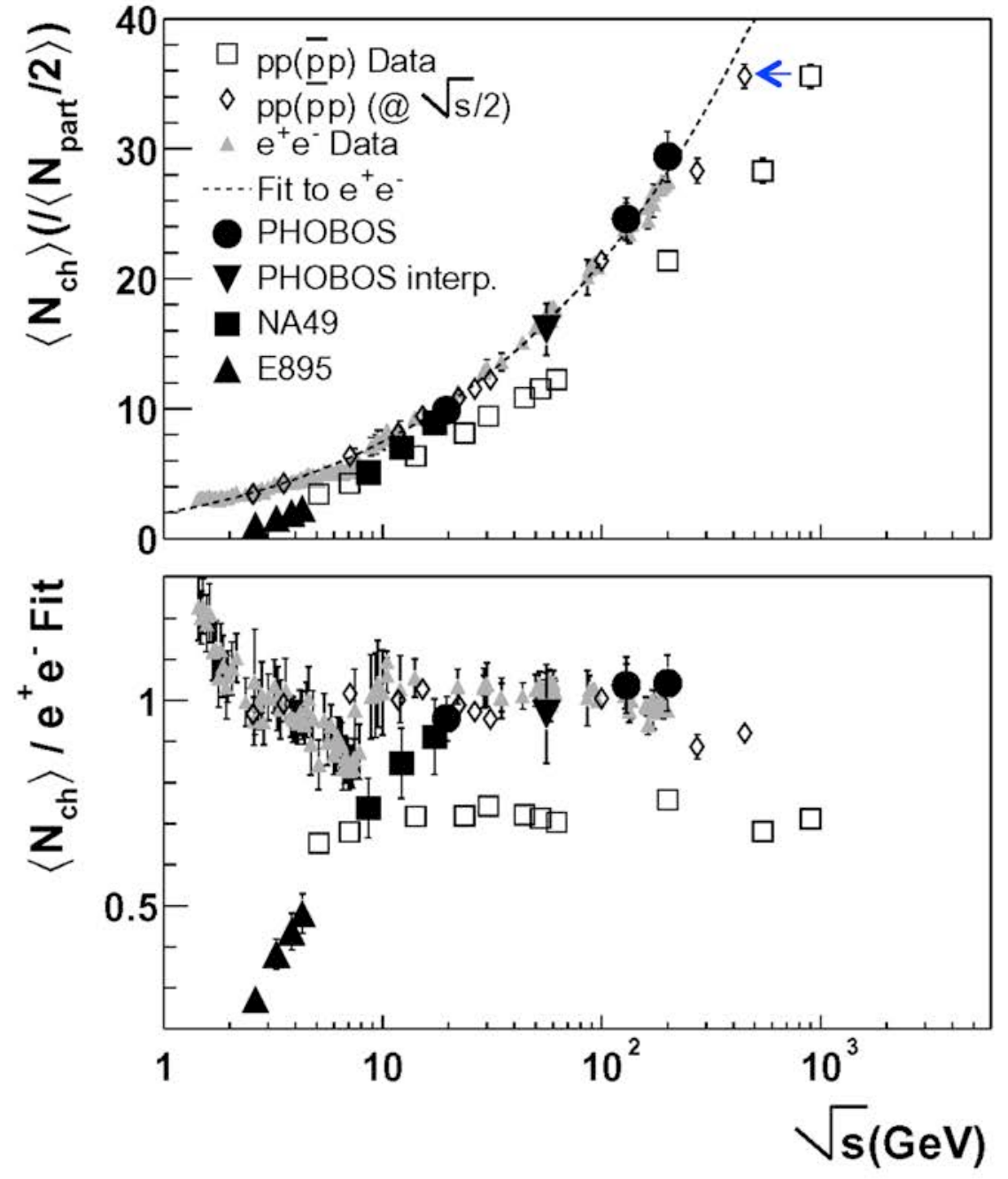}
\end{center}\vspace*{-1.7pc}
\caption[]{a) (left) Charged particle multiplicity density in rapidity, $dN_{\rm ch}/d\eta$, as a function of centrality in d+Au collisions at $\sqrt{s_{NN}}=200$ GeV~\cite{PHOBOSPRC72}. b)~(right) Total charged multiplicity per nucleon pair in p-p and A+A collisions as a function of c.m. energy $\sqrt{s}$ compared to $e^+ + e^-$ collisions~\cite{PHOBOSPRC74}.
\label{fig:PHOBOSdAAA}}\vspace*{-1.8pc}
\end{figure}

Returning to the soft physics of multiplicity and $E_T$ distributions, the PHOBOS experiment at RHIC, with a large pseudo-rapidity acceptance $-5<\eta<+5$ over the full azimuth had presented an instructive measurement of the charged multiplicity density, $dN_{\rm ch}/d\eta$ from the first d+Au run in 2003 (Fig.~\ref{fig:PHOBOSdAAA}a)~\cite{PHOBOSPRC72}.
For peripheral collisions, the distribution is symmetric around mid-rapidity, as in p-p collisions. However, with increasing centrality, the multiplicity increases over the whole $\eta$ range but with a larger increase at negative $\eta$ (the Au rapidity) such that the peak of the distribution steadily shifts in the direction of the Au nucleus. 

These features which are similar to what was first observed in fixed target p+A experiments at $\sqrt{s_{NN}}\sim 19.4$ GeV could be explained (c. 1976) by a simple model, the Wounded Nucleon Model (WNM)~\cite{WNM}. From relativity and quantum mechanics the only thing that can happen to a relativistic nucleon when it interacts with another nucleon in a nucleus is to become an excited nucleon with the same energy but reduced longitudinal momentum (rapidity). It remains in that state inside the nucleus because the uncertainty principle and time dilation prevent it from fragmenting into particles until it is well outside the nucleus. 
If one makes the further assumptions that an excited nucleon interacts with the same 
cross section as an unexcited nucleon and that the successive collisions of the excited nucleon do not affect the excited state or its eventual fragmentation products, this leads to the conclusion that the elementary process for particle 
production in nuclear collisions is the excited nucleon, and to the prediction 
that the multiplicity in nuclear interactions should be proportional to 
the total number of projectile and target participants (Wounded Nucleons)~\cite{WNM}, rather than to the 
total number of collisions. 

Another interesting effect observed by PHOBOS~\cite{PHOBOSPRC74} is that the ``leading particle effect'' in p-p collisions, as discovered by Zichichi and collaborators~\cite{BasilePLB95}, in which the total multiplicity at c.m. energy $\sqrt{s_{\rm pp}}$ is equal to that in $e^+ e^-$ collisions at $\sqrt{s_{\rm ee}}=\sqrt{s_{\rm pp}}/2$ (the ``effective energy'') because the leading protons carry away half the p-p c.m. energy, is absent in A+A collisions (Fig.~\ref{fig:PHOBOSdAAA}b). This observation seems to contradict the WNM, in which the key assumption is that what counts is whether or not a nucleon was struck, not how many times it was struck.
\begin{figure}[!t]
   \begin{center}
\includegraphics[width=0.44\textwidth]{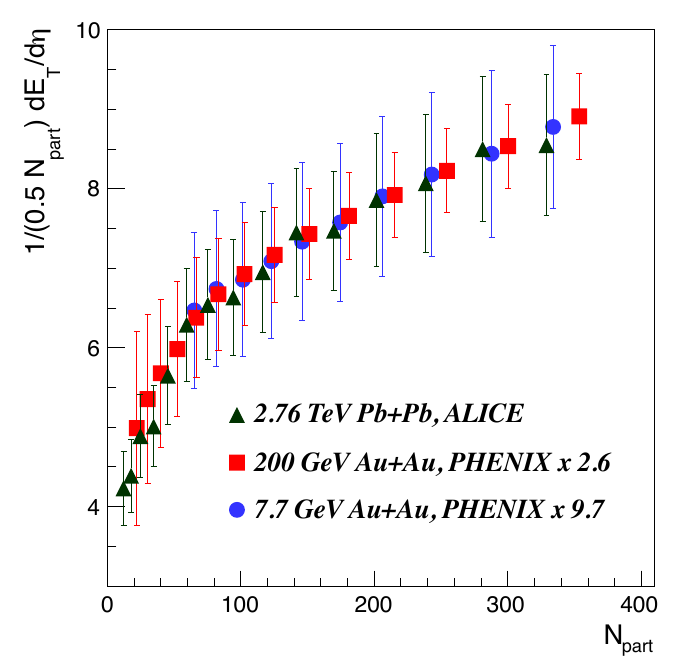}\hspace*{0.4pc} 
\includegraphics[width=0.55\textwidth]{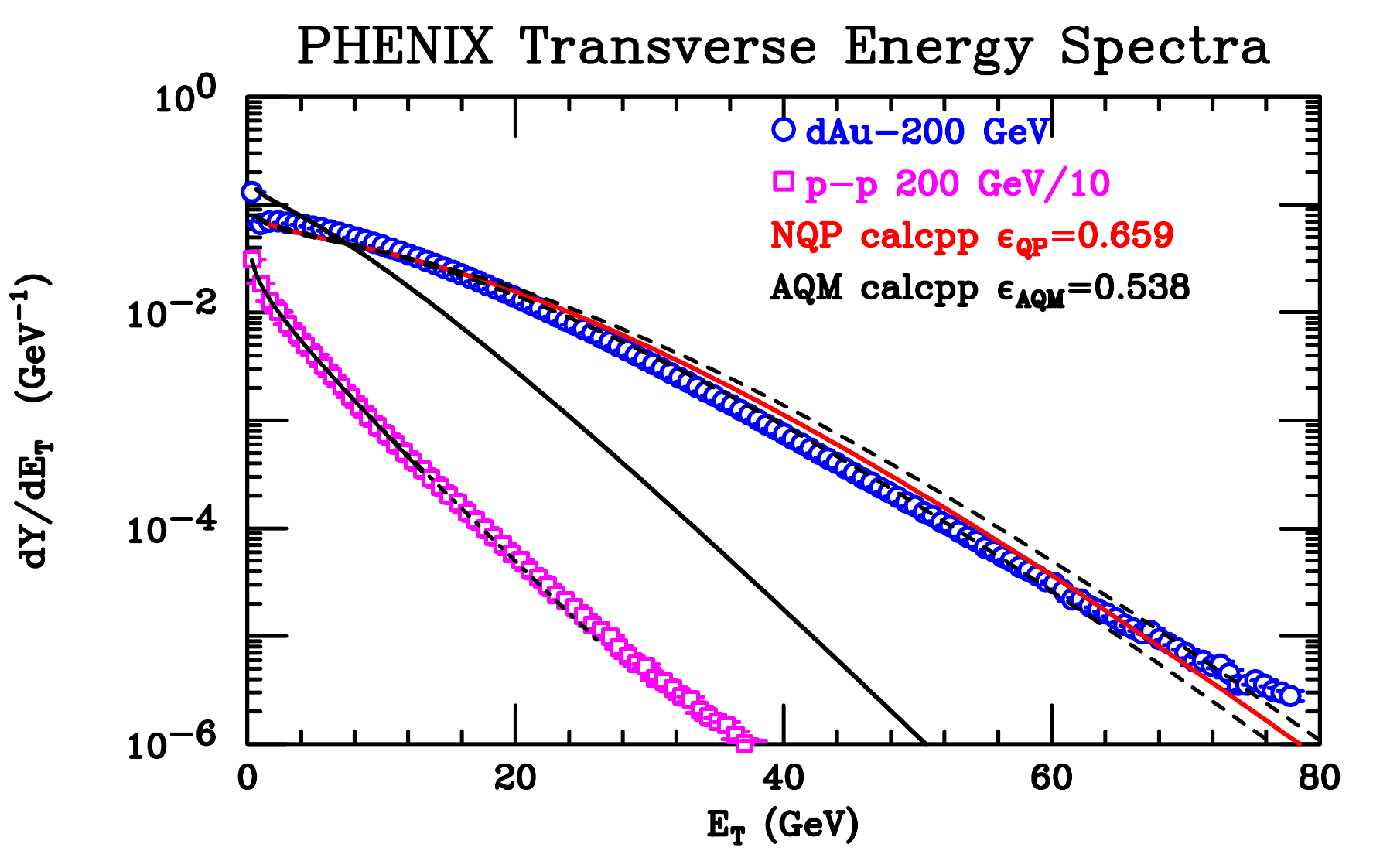}
\end{center}\vspace*{-1.7pc}
\caption[]{a) (left) $dE_T^{AA}/d\eta/(0.5 \mean{N_{\rm part}})$ vs. $\mean{N_{\rm part}}$ in Au+Au and Pb+Pb collisions from $\sqrt{s_{NN}}=0.0077$ to 2.76 TeV. b) (right) PHENIX preliminary measurement of $E_T$ distributions for p+p and d+Au at $\sqrt{s_{NN}}=200$ GeV  with calculations of the d+Au spectrum based on the AQM (color-strings) and the number of constituent-quark participants (NQP).
\label{fig:NQP}}\vspace*{-0.8pc}
\end{figure}

In fact, the WNM fails badly at mid-rapidity for both $dN_{ch}/d\eta$ and $dE_T/d\eta$ as shown by a plot of $dE_T/d\eta/(N_{\rm part}/2)$ vs. $N_{\rm part}$ from PHENIX which should be constant if the WNM were true (Fig.~\ref{fig:NQP}a). 
The fact that the scaled evolution with centrality is the same from $\sqrt{s_{NN}}=7.7$ GeV to 2.76 TeV indicates that the dominant effect is the nuclear geometry of the A+A collision. It has been shown that the evolution in Fig.~\ref{fig:NQP}a can be explained by a nuclear geometry based on the number of constituent-quark participants, the NQP model~\cite{VoloshinNQP,NouicerNQP}.
Thus the shape of the data in Fig.~\ref{fig:NQP}a is simply the number of constituent-quark participants/nucleon participant, ${N_{\rm qp}}/N_{\rm part}$.  

For symmetric systems such as Au+Au, the NQP model is identical to another model from the 1970's, the Additive Quark Model (AQM)~\cite{AQM}.  The AQM is actually a model of particle production by color-strings in which only one color-string can be attached to a constituent-quark participant. Thus, for asymmetric systems such as d+Au, the maximum number of color-strings is limited to the number of constituent-quarks in the lighter nucleus, or six for d+Au, while the NQP allows all the quark participants in both nuclei to emit particles. Fig.~\ref{fig:NQP}b shows that the NQP model gives the correct $E_T$ distribution in d+Au, while the AQM has a factor of 1.7 less $E_T$ emission due to the restriction on the number of effective constituent-quarks in the larger nucleus. The positions of the three-constituent quarks are generated about the position of each nucleon, in a standard Glauber calculation, according to the measured charge distribution of the proton, which gives a physical basis for the ``proton size fluctuations'' recently discussed at LHC~\cite{ATLAS-Gribov}. 

Constituent-quark-participants might also explain the increase of the ``effective energy'' in A+A collisions compared to p-p collisions as discussed above (Fig.~\ref{fig:PHOBOSdAAA}b). The $\mean{N_{\rm qp}/N_{\rm part}}$ is 1.5 for a p-p collision but rises to 2.3-2.7 for more central (0-50\%) A+A collisions. Thus the ``effective energy'' for particle production increases due to an increase in the number of (constituent-quark) participants, not because of additional collisions of a given participant. This preserves the assumption in these ``extreme-independent'' participant models that successive collisions of a participant do not increase its particle emission.  

The most surprising soft-physics result in p+Pb and d+Au physics this year concerns what looks very much like collective flow observed in these small systems, where no (or negligible) medium or collective effect was expected.  
         \begin{figure}[!h]
   \begin{center}
\includegraphics[width=0.95\textwidth]{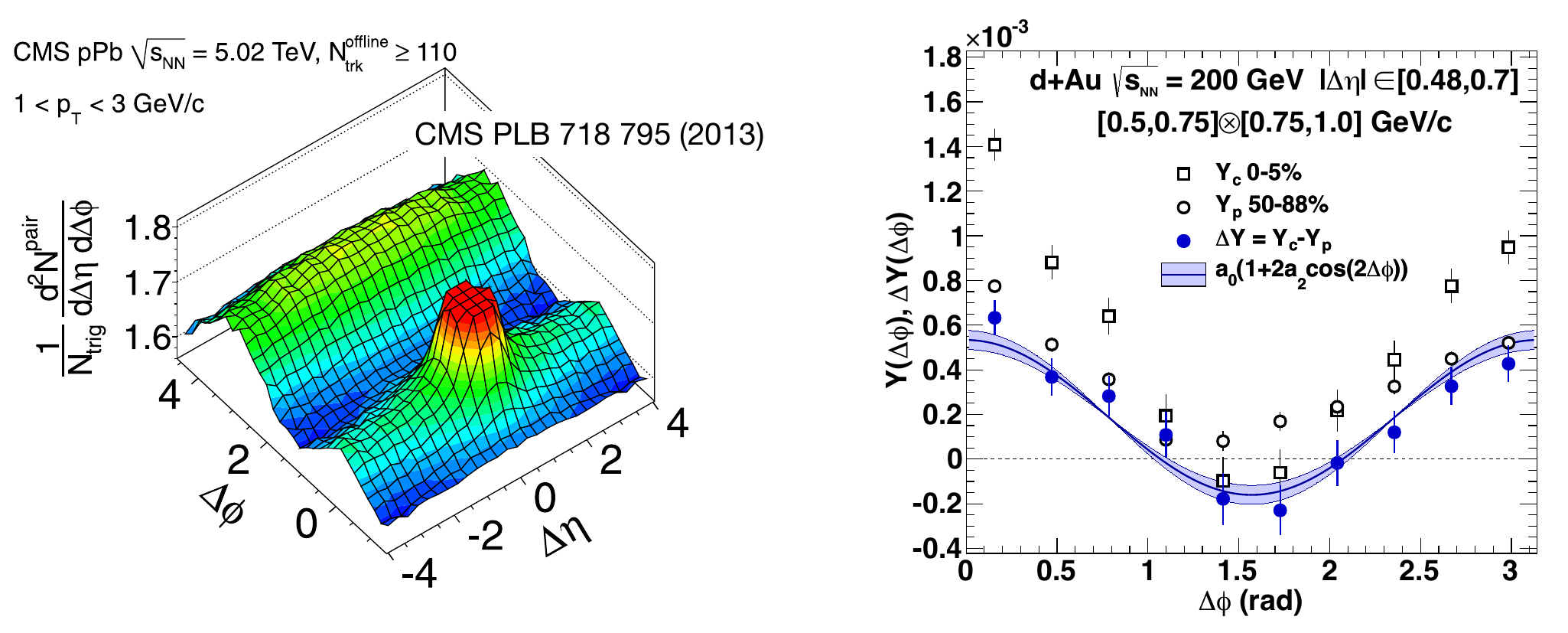}
\end{center}\vspace*{-1.7pc}
\caption[]{a) (left) CMS ridge in two-particle correlations~\cite{CMSPLB718}. b) (right) PHENIX two-particle azimuthal correlations in d+Au at RHIC~\cite{PXv2dAu}.
\label{fig:RidgedAu}}\vspace*{-0.8pc}
\end{figure}
Fig.~\ref{fig:RidgedAu}a shows a LEGO plot of $\Delta\eta$, $\Delta\phi$, the difference in polar and azimuthal angles from correlations of two particles with $1<p_T<3$ GeV/c in p+Pb by CMS at $\sqrt{s_{NN}}=5.02$ TeV~\cite{CMSPLB718}. A clear $1+2v_2\cos 2\Delta\phi$ modulation of the distribution independent of $\Delta\eta$ is observed, called the `ridge' in Au+Au collisions where the modulations $v_2$, $v_3$, \ldots $v_n$ are attributed to collective flow of the QGP medium. 
At RHIC, PHENIX confirmed this result in d+Au collisions at $\sqrt{s_{NN}}=200$ GeV (Fig.~\ref{fig:RidgedAu}b)~\cite{PXv2dAu}. In order to remove any $v_2$ effect due to two-particle correlations of hard-scattering, lower $p_T$ triggers were used as well as cuts in $\Delta\eta$ to remove the same-side peak. Also, since there is no suppression of hard-scattering in p+Pb or d+Au collisions (recall Fig.~\ref{fig:RAAdAuAuAu}), the conditional two-particle yield from di-jets is independent of centrality. Thus, any residual hard-scattering effect was removed by subtracting the peripheral (50-88\%) from the central (0-5\%) measurement which revealed the beautiful $\cos 2\Delta\phi$ curve characteristic of elliptical flow shown in Fig.~\ref{fig:RidgedAu}b.  

\begin{figure}[!h]
   \begin{center}
\raisebox{0.6pc}{\includegraphics[width=0.51\textwidth]{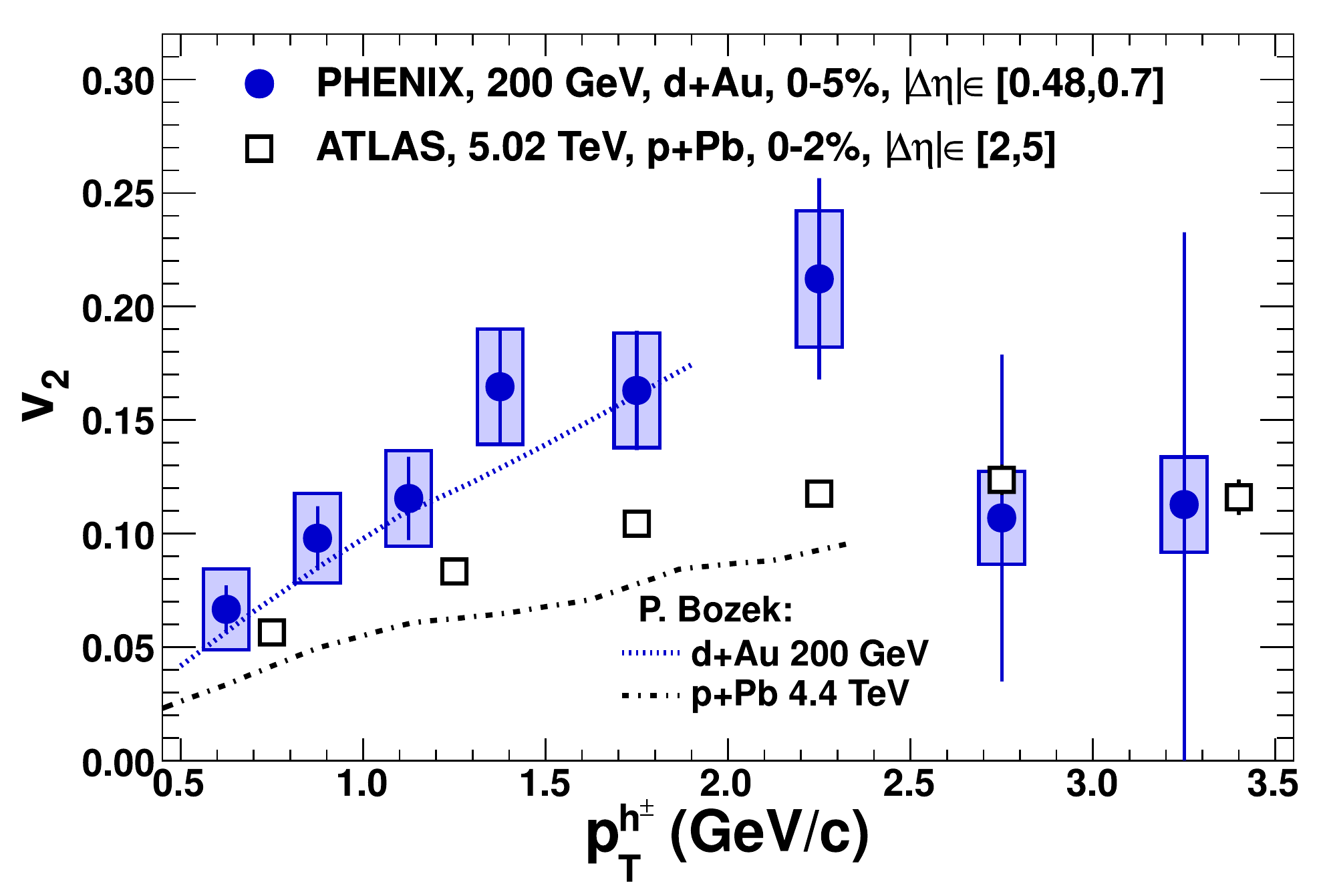}}
{\includegraphics[width=0.47\textwidth]{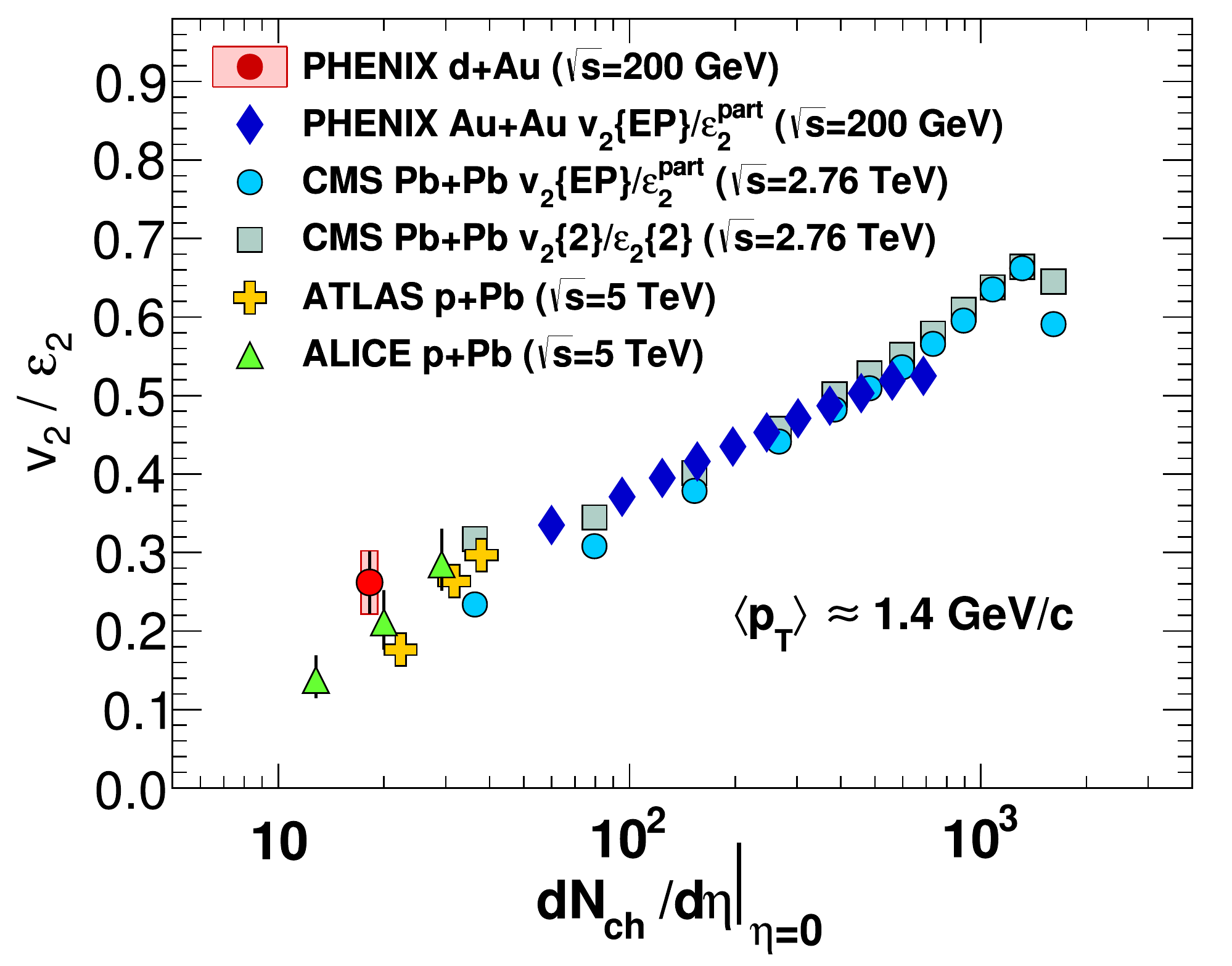}}
\end{center}\vspace*{-1.7pc}
\caption[]{a) (left) Comparison of $v_2$ vs. $p_T$ in d+Au at RHIC to p+Pb at LHC~\cite{PXv2dAu}. b) (right) Compilation of $v_2/\varepsilon$ vs. $dN_{\rm ch}/d\eta$ at $p_T=1.4$ GeV/c in d+Au, p+Pb, Au+Au and Pb+Pb collisions at RHIC and LHC~\cite{PXv2dAu}. 
\label{fig:PXv2dAu}}\vspace*{-0.8pc}
\end{figure}

Figure~\ref{fig:PXv2dAu}a compares the $v_2$ measurements vs $p_T$ from d+Au at $\sqrt{s_{NN}}=200$ GeV and p+Pb at 5.02 TeV. The larger values from the d+Au results are thought to be due to the larger eccentricity ($\varepsilon$) of the two-nucleon deuteron compared to the single nucleon proton. In fact, the values of $v_2/\varepsilon$ from d+Au and p+Pb are consistent with the dependence of $v_2/\varepsilon$ on $dN_{\rm ch}/d\eta$ (Fig.~\ref{fig:PXv2dAu}b) observed in Au+Au and Pb+Pb collisions, which was taken as proof of collective flow from hydrodynamics. 

\begin{figure}[!h]
   \begin{center}
\includegraphics[width=0.95\textwidth]{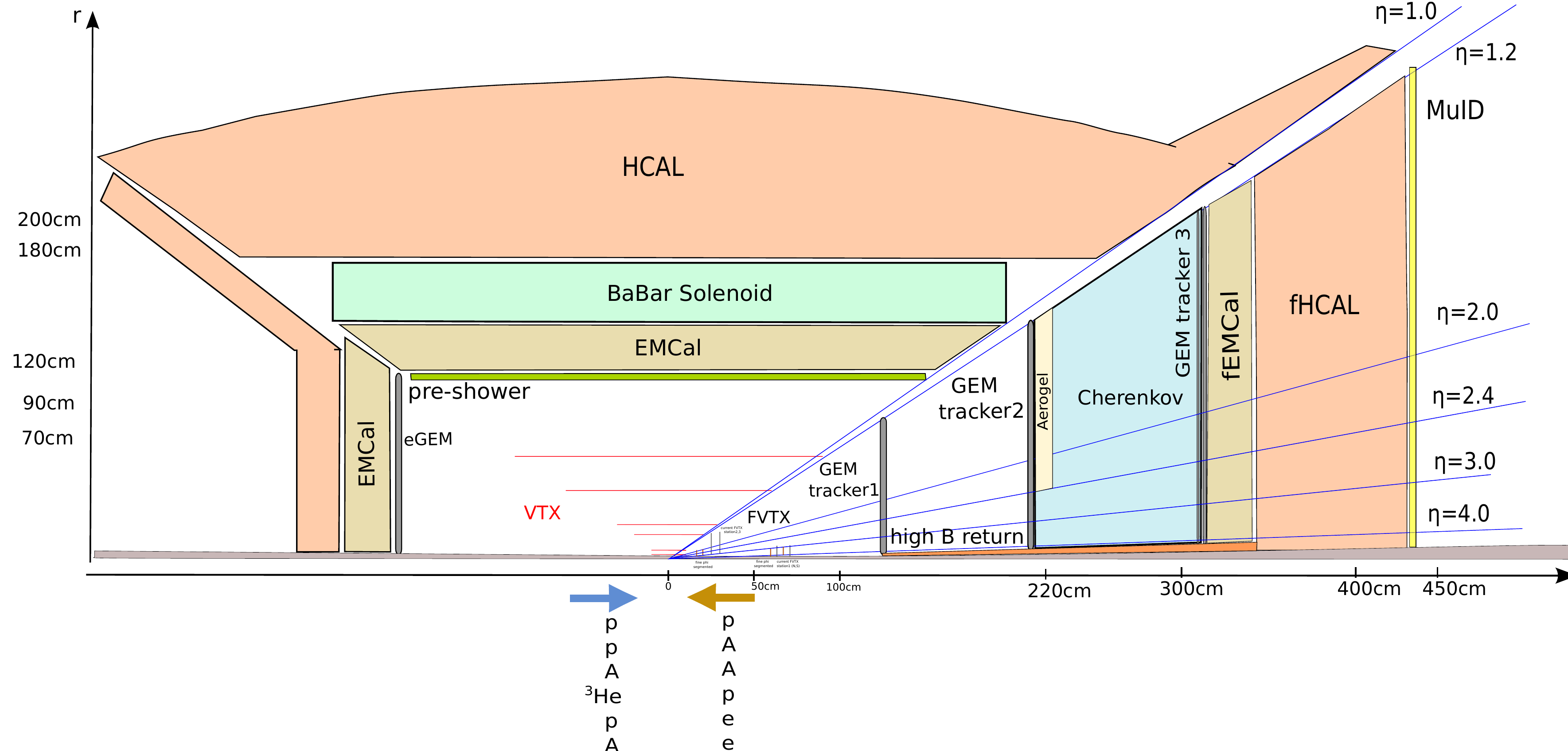}
\end{center}\vspace*{-1.7pc}
\caption[]{a) sPHENIX concept with forward detector~\cite{sPHENIX2013}.
\label{fig:ePHENIX}}\vspace*{-0.8pc}
\end{figure}

These new results again underscore the importance of p-p and p+A comparison data to understand the observations in RHI collisions, where the detailed physics of the QGP is far from understood. To this end, PHENIX has proposed a new more conventional collider detector, s(uper)PHENIX, based on a thin-coil superconducting solenoid, to concentrate on hard-scattering and jets. This would replace the very successful but 15 year old special purpose small aperture two-arm spectrometer designed to measure $J/\Psi\rightarrow e^+ + e^-$ down to zero $p_T$ at mid-rapidity, the original expected signal for deconfinement, as well as identified particles such as single $e^{\pm}$ from heavy quark decay, $\pi^0$, $\eta$ and other hadrons (recall Fig.~\ref{fig:Tshirt}b) that could cause background to the $J/\Psi$ but which turned out to be valuable probes of the QGP. This year sPHENIX got a big boost by acquiring the (made in Italy) BABAR solenoid magnet from SLAC which became available when the B-factory in Italy was unfortunately cancelled. Conceptual design of the new experiment is well underway, with mid-rapidity, forward and eRHIC capability~\cite{sPHENIX2013} (Fig.~\ref{fig:ePHENIX}). New collaborators would be most welcome.   

\end{document}